\documentclass[10pt,journal,compsoc]{IEEEtran}
\IEEEoverridecommandlockouts
\usepackage{enumitem}
\usepackage{cite}
\usepackage{amsmath,amssymb,amsfonts}
\usepackage{graphicx}
\usepackage{textcomp}
\usepackage{xcolor}
\usepackage [english]{babel}
\usepackage [autostyle, english = american]{csquotes}
\MakeOuterQuote{"}
\usepackage{graphicx}
\usepackage{float}
\usepackage{graphics} 
\usepackage{epsfig} 
\usepackage{booktabs}
\usepackage{dsfont}
\usepackage{subfigure}
\usepackage{tabularx}
\usepackage{tabularx}
\usepackage{pdflscape}
\usepackage{bbding}
\usepackage{wasysym}
\usepackage{amssymb}
\usepackage{pifont}
\usepackage{amssymb}
\usepackage{afterpage}
\usepackage{rotating}
\usepackage{lscape}
\usepackage{hyperref}
\hypersetup{bookmarks=false}
\usepackage{amsmath}
\graphicspath{ {images/} }
\usepackage{amsmath}
\usepackage{mathtools}

\DeclarePairedDelimiter\floor{\lfloor}{\rfloor}

\usepackage{lineno,hyperref}
\usepackage{algorithm}
\usepackage{float}
\usepackage{algpseudocode}
\usepackage{mathtools}
\algdef{SE}{Begin}{End}{\textbf{Begin}}{\textbf{End}}
\algdef{SE}[PROCEDURE]{Procedure}{EndProcedure}%
   [2]{\algorithmicprocedure\ \textproc{#1}\ifthenelse{\equal{#2}{}}{}{(#2)}}%
   {\algorithmicend\ \algorithmicprocedure}%
\algdef{SE}[DOWHILE]{Do}{doWhile}{\algorithmicdo}[1]{\algorithmicwhile\ #1}%
\newlength\myindent
\usepackage{xcolor}
\newcommand{\blue}[1]{\textcolor{black}{#1}}

\def\BibTeX{{\rm B\kern-.05em{\sc i\kern-.025em b}\kern-.08em
    T\kern-.1667em\lower.7ex\hbox{E}\kern-.125emX}}
    
\makeatletter
\patchcmd{\@maketitle}
  {\addvspace{0.5\baselineskip}\egroup}
  {\addvspace{-1\baselineskip}\egroup}
  {}
  {}
\makeatother

\begin{document}

\title{START: Straggler Prediction and Mitigation for Cloud Computing Environments using Encoder LSTM Networks}

\author{
        Shreshth~Tuli$^1$,
        Sukhpal~S.~Gill$^2$,
        Peter~Garraghan$^3$,
        Rajkumar~Buyya$^4$,
        Giuliano~Casale$^1$,
    and~Nicholas~R.~Jennings$^{1, 5}$
\IEEEcompsocitemizethanks{
\IEEEcompsocthanksitem S. Tuli, G. Casale and N. R. Jennings are with the $^1$Department of Computing, Imperial College London, United Kingdom.\protect
\IEEEcompsocthanksitem N. R. Jennings is also with $^5$Loughborough University, United Kingdom.\protect
\IEEEcompsocthanksitem S.S. Gill is with the $^2$School of Electronic Engineering and Computer Science, Queen Mary University of London, United Kingdom.\protect
\IEEEcompsocthanksitem P. Garraghan is with the $^3$School of Computing and Communications, Lancaster University, United Kingdom.\protect
\IEEEcompsocthanksitem R. Buyya is with the $^4$Cloud Computing and Distributed Systems (CLOUDS) Laboratory, University of Melbourne, Australia.\protect\\
E-mail: s.tuli20@imperial.ac.uk, s.s.gill@qmul.ac.uk, p.garraghan@lancaster.ac.uk, rbuyya@unimelb.edu.au, g.casale@imperial.ac.uk, n.r.jennings@lboro.ac.uk.}
\thanks{Manuscript received ---; revised ---.}}

\markboth{IEEE Transactions on Services Computing}%
{Tuli \MakeLowercase{\textit{et al.}}: Straggler prediction and mitigation using Encoder LSTM}

\IEEEtitleabstractindextext{%
\begin{abstract}
Modern large-scale computing systems distribute jobs into multiple smaller tasks which execute in parallel to accelerate job completion rates and reduce energy consumption. However, a common performance problem in such systems is dealing with straggler tasks that are slow running instances that increase the overall response time. Such tasks can significantly impact the system's Quality of Service (QoS) and the Service Level Agreements (SLA). To combat this issue, there is a need for automatic straggler detection and mitigation mechanisms that execute jobs without violating the SLA. Prior work typically builds reactive models that focus first on detection and then mitigation of straggler tasks, which leads to delays. Other works use prediction based proactive mechanisms, but ignore heterogeneous host or volatile task characteristics. In this paper, we propose a Straggler Prediction and Mitigation Technique (START) that is able to predict which tasks might be stragglers and dynamically adapt scheduling to achieve lower response times. Our technique analyzes all tasks and hosts based on compute and network resource consumption using an Encoder Long-Short-Term-Memory (LSTM) network. The output of this network is then used to predict and mitigate expected straggler tasks. This reduces the SLA violation rate and execution time without compromising QoS. Specifically, we use the CloudSim toolkit to simulate START in a cloud environment and compare it with state-of-the-art techniques (IGRU-SD, SGC, Dolly, GRASS, NearestFit and Wrangler) in terms of QoS parameters such as energy consumption, execution time, resource contention, CPU utilization and SLA violation rate. Experiments show that START reduces execution time, resource contention, energy and SLA violations by 13\%, 11\%, 16\% and 19\%, respectively, compared to the state-of-the-art approaches.
\end{abstract}

}

\maketitle
\IEEEdisplaynontitleabstractindextext
\IEEEpeerreviewmaketitle

\section{Introduction}
\label{sec:intro}

Emerging applications of Cloud Data-Centers (CDCs) in domains such as healthcare, agriculture, smart cities, weather forecasting and traffic management produce large volumes of data, which is transferred among different devices using various kinds of communication modes~\cite{gill2020tails}. Due to this continuous increase in data volume and velocity,  large-scale computing systems may be utilized~\cite{xu2016optimization, liaqat2019characterizing, mustafa2019sla}, which exacerbates the need for scalable, automated scheduling and intelligent task placement methods. This work focuses on this problem by studying, in particular, strategies to mitigate straggler tasks. Stragglers are tasks within a job that take much longer to execute than other tasks and can cause a significant increase in response time due to the need for synchronizing the outputs of the tasks. Their presence can lead to the so-called Long Tail Problem~\cite{wang2015using}.  

More precisely, the Long Tail Problem occurs when the completion time of a particular job is significantly affected by a small number of straggler tasks in a negative way. Task stragglers can occur within any highly parallelized system that processes jobs consisting of multiple tasks. Google’s MapReduce framework~\cite{coppa2015data} or the Hadoop framework~\cite{eldawy2015spatialhadoop} are examples of such systems, where solutions for straggler prevention are common~\cite{gill2020tails, ananthanarayanan2014grass, bitar2020stochastic}. Both MapReduce and Hadoop allow for scalability of the system to vast clusters of commodity servers. The parallel execution of tasks increases the speed of execution and handles the failures automatically without human intervention following the principles of IBM’s autonomic model~\cite{gill2019holistic, kosta2012thinkair}. However, stragglers can still occur because of software/hardware faults as autonomic models are often slow in handling failures and can result in long down-times in resource-constrained devices~\cite{gill2020tails}. These lead to unexpected delays in task execution due to resource unavailability or data loss and cause such tasks to hog resources which in non-preemptive execution leads to higher response times. Thus, efficient techniques are required to mitigate stragglers to prevent high response times and SLA violations. We now discuss what types of failures lead to stragglers tasks.

There are two types of failures that can occur during the execution of jobs: task failures and node failures. The former occurs when a specific task within a job fails, due to diverse sources of software and hardware faults~\cite{lindsay2019prism}. The latter occurs when one of the resources of a specific node, which executes the job’s task, fails~\cite{gill2020tails}. This can be caused by a myriad of possible OS or hardware level faults. As an example of straggler mitigation techniques, MapReduce attempts to mitigate task failures by relaunching the task once it fails~\cite{garraghan2018emergent}. In terms of a node failure, MapReduce re-executes all the tasks that were originally scheduled to be executed on that node. In terms of node failures, when the performance of a node degrades, either due to an OS or hardware fault or the node completely fails, a specific task’s (straggler) execution time can be bloated, causing any other tasks that depend on it to wait for its completion~\cite{wang2014efficient}. \blue{At the job level, for the job to be considered complete, all the tasks comprising the job must finish. If a straggling task prevents other sibling tasks from successfully completing, the job will not be complete until all the straggler tasks are complete~\cite{kumar2014comprehensive}. Furthermore, straggler tasks can keep other tasks dependent on their output waiting and hence consume additional resources, further impacting the performance of the computing system.}

Stragglers not only affect performance but also deployment costs. Popular cloud service providers such as Amazon, Google, Netflix and Apple face the challenge of straggler tasks leading to delayed response or resource wastage. This requires avoidable scaling-up of the cloud infrastructures, which in turn increase the deployment costs~\cite{aktas2017effective, wang2014efficient}. 
The high latency episodes called “tail-tolerant” or “latency-tail-tolerant”, also affect the performance of cloud services~\cite{yadwadkar2014wrangler}.  Latency tail-tolerant jobs reduce resource utilization and increase energy consumption. Characterization studies such as~\cite{gill2020tails, xu2016optimization, wang2015using, coppa2015data, farhat2015stochastic, gill2019holistic, lindsay2019prism}, show that resource contention is the main reason for stragglers, occurring when different jobs are waiting for shared resources. Different applications executing on different nodes may also contend for shared global resources~\cite{yadwadkar2014wrangler}. 


Prior work~\cite{zaharia2012resilient, ananthanarayanan2013effective} focuses on solving the problem of straggler tasks by detecting and mitigating which tasks are stragglers only after the jobs are executed. Straggler mitigation refers to the prevention of any impact of straggler tasks on QoS or SLA. This not only requires continuous computation resources, but these monitoring tasks themselves can be so data-intensive that they can themselves lead to resource contention, delays and prevent scalability of the system~\cite{gill2019transformative}. However, modern technologies like deep learning allow us to build scalable models to not only detect, but predict beforehand, which tasks might be straggler and run mitigation algorithms to save time and improve QoS. Here, straggler prediction means the prediction of straggler tasks before execution. In particular,~\cite{lu2019gru, fang2012rpps} use deep learning based solutions to predict straggler tasks and efficiently manage them. 

Deep learning based straggler prediction methods face large prediction errors due to two major problems. First, these models ignore the underlying distribution of task execution times which is crucial to determine straggler tasks~\cite{gill2020tails, xu2016optimization}. Specifically, diversity in task execution times leads to the presence of tasks with extremely high or low execution times. This makes the state space of the neural network very large when modelling the distribution of task response times and hence it is often omitted in practical approaches~\cite{lu2019gru, fang2012rpps}. Second, these approaches ignore the heterogeneous host capabilities, which can also lead to poor scheduling or mitigation decisions \cite{gill2019transformative}. Therefore, a new method is required which can both proactively predict straggler tasks and efficiently mitigate them. As an example of a heterogeneous execution environment, fog-cloud environments leverage resource capabilities from both edge devices and cloud nodes~\cite{gill2019transformative}. This leads to high diversity in the computational resources among host devices in the same environment. This host heterogeneity impacts the response time as scheduling in a constrained device may significantly increase its response time.

These issues motivate us to develop a novel online SLA-aware \textbf{ST}r\textbf{A}ggler P\textbf{R}ediction and Mi\textbf{T}igation (START) technique. START uses a machine learning model in tandem with an underlying distribution or task response time for automatic and accurate straggler prediction. To allow mapping of heterogeneous environments, encoder networks have shown to be a promising solution~\cite{tuli2019healthfog}. Moreover, prior works also show that in dynamic environments, Long-Short-Term-Memory (LSTM) based neural networks help to adapt to environment changes~\cite{gill2020thermosim}. Hence, we use an Encoder-LSTM network to analyze the state of a cloud environment. \blue{Here, the state of the cloud setup is characterized as a set of host and task parameters like SLA, CPU, RAM, Disk and bandwidth consumption. These parameters are motivated by prior work~\cite{tuli2021cosco}.} Further, as prior work has shown that response times of tasks in large-scale cloud setups follow a Pareto distribution~\cite{gill2020tails}, we use the Encoder-LSTM network to predict this distribution in advance to alleviate the straggler problem proactively.

START also uses speculation and rerun-based approaches for Straggler Mitigation during the execution of jobs. Prediction allows early mitigation, reducing the SLA violation rate and execution time and maintaining QoS at the required level. Our performance evaluation is carried out using CloudSim 5.0~\cite{calheiros2011cloudsim} and compares our technique with well-known existing techniques (SGC~\cite{bitar2020stochastic}, Dolly~\cite{ananthanarayanan2013effective}, GRASS~\cite{ananthanarayanan2014grass}, NearestFit~\cite{coppa2015data}, Wrangler~\cite{yadwadkar2014wrangler}, and IGRU-SD~\cite{lu2019gru}) in terms of QoS parameters such as energy consumption, execution time, resource contention, CPU utilization and SLA violation rate. Experimental results demonstrate that START gives lower execution time and SLA violations than existing techniques, also offering low computational overhead.

The rest of the paper is structured as follows. Section \ref{sec:relwork} presents related work. Section \ref{sec:model} details START. Sections \ref{sec:setup} and \ref{sec:exp} describe the evaluation setup and experimental results. Finally, Section \ref{sec:conclusion} concludes and outlines future research directions. 

\section{Related Work}
\label{sec:relwork}

\begin{table*}[t]
    \centering
    \caption{Comparison of existing models with START}
    \resizebox{\textwidth}{!}{
    \begin{tabular}{@{}lccccccc@{}}
    \toprule 
    Technique & Straggler Detection & Straggler Mitigation & Proactive Mechanism  & Straggler Prediction & Impact on QoS and Utilization & Dynamic & Heterogeneous Environment \tabularnewline
    \midrule
    \multicolumn{8}{c}{Detection Only Methods}
    \tabularnewline
    \midrule
    NearestFit~\cite{coppa2015data} & \checkmark &  &  &  &  & \checkmark & \tabularnewline
    SMT~\cite{ouyang2016straggler} & \checkmark &  &  &  & \checkmark &  & \tabularnewline
    SMA~\cite{wang2014efficient} & \checkmark &  &  &  &  &  & \tabularnewline
    RDD~\cite{zaharia2012resilient} & \checkmark &  &  &  &  &  & \tabularnewline
    \midrule
    \multicolumn{8}{c}{Mitigation Only Methods}
    \tabularnewline
    \midrule
     LATE~\cite{zaharia2008improving} &  & \checkmark &  &  & \checkmark &  & \checkmark\tabularnewline
    Dolly~\cite{ananthanarayanan2013effective}  &  & \checkmark & \checkmark &  &  &  & \tabularnewline
    GRASS~\cite{ananthanarayanan2014grass} &  & \checkmark & \checkmark &  & \checkmark &  & \tabularnewline
    Dolly~\cite{ananthanarayanan2013effective}  &  & \checkmark & \checkmark &  &  &  & \tabularnewline
    GRASS~\cite{ananthanarayanan2014grass} &  & \checkmark & \checkmark &  & \checkmark &  & \tabularnewline
    Wrangler~\cite{yadwadkar2014wrangler} &  & \checkmark &  &  &  & \checkmark & \tabularnewline
    \midrule
    \multicolumn{8}{c}{Prediction based Mitigation Methods}
    \tabularnewline
    \midrule
    SGC~\cite{bitar2020stochastic} & \checkmark & \checkmark & \checkmark & \checkmark & \checkmark &  & \checkmark\tabularnewline
    IGRU-SD~\cite{lu2019gru} & \checkmark & \checkmark & \checkmark & \checkmark &  & \checkmark & \tabularnewline
    \textbf{START (this work)} & \checkmark & \checkmark & \checkmark & \checkmark & \checkmark & \checkmark & \checkmark\tabularnewline
    \bottomrule 
    \end{tabular}
    }
    \label{tab:related-work}
\end{table*}


Existing straggler analysis and mitigation techniques can be mainly divided into two main categories: detection and mitigation~\cite{gill2020tails, xu2016optimization}. The former  primarily identify stragglers from utilization metrics and traces from a job execution environment like a CDC. Most of these techniques leverage offline analytics and real-time monitoring methods. Examples of such techniques include NearestFit~\cite{coppa2015data} and SMT~\cite{ouyang2016straggler}. Within this category, other techniques use prediction models to a-priori determine the set of tasks in a job that might be stragglers. Examples include RPPS~\cite{fang2012rpps} and IGRU-SD~\cite{lu2019gru}. When considering mitigation, approaches either avoid straggler tasks or prevent high response times by methods such as re-scheduling, balancing load or running job replicas (clones). Examples of such strategies include Dolly~\cite{ananthanarayanan2013effective}, GRASS~\cite{ananthanarayanan2014grass}, LATE~\cite{zaharia2008improving} and Wrangler~\cite{yadwadkar2014wrangler}. Table~\ref{tab:related-work} summarizes the comparison of START with prior approaches. The table shows which works use straggler prediction, mitigation and/or detection. Further, \textit{proactive mechanism} shows if methods use prediction data to proactively mitigate straggler tasks or wait till completion of other tasks. \textit{Impact on QoS and Utilization} shows whether these methods utilize QoS and host utilization metrics as feedback to improve prediction or mitigation performance. \textit{Dynamic} refers to whether these methods are able to adapt to changing host/task characteristics. \textit{Heterogeneous environment} refers to whether a method assumes resources to have the same computational characteristics.

\textbf{Straggler Detection.}
The NearestFit strategy aims at improving the performance of distributed computing systems by resolving data skewness and detecting straggler tasks or unbalanced load. Through this model, \cite{coppa2015data} proposes a fully-online nearest neighbor regression method that uses statistical techniques to profile the tasks running in the system. This model gathers profiles using efficient data streaming algorithms and acts as a progress indicator and it is therefore suited to applications with long run times. Even though this indicator is able to profile complex and large-scale systems, it is not suitable for heterogeneous resource types as it does not differentiate hosts on the basis of computational capacities. Further, it does not take into account task failures or load on each host. 

\textbf{Straggler Prediction.}
The work in~\cite{fang2012rpps} proposes a resource prediction and provisioning scheme (RPPS) using the Autoregressive Integrated Moving Average (ARIMA) model, which is a statistical model for the prediction of future workload characteristics of various tasks running in a CDC. The work in~\cite{lu2019gru} very recently proposed a technique called Improved Gated Recurrent Unit with Stragglers Detection (IGRU-SD) to predict average resource requests over time. They use this prediction scheme to then run detection algorithms for predicting which tasks might be a straggler. However, they do not consider host heterogeneity, nor do they consider the underlying task distribution, both of which are crucial for predicting if a task is likely to become a straggler.

\textbf{Straggler Mitigation.}
The work in~\cite{ananthanarayanan2013effective} explores straggler mitigation techniques and proposes, Dolly, a speculative execution-based approach that launches multiple clones of expected straggler tasks and takes the results of the clone, which finishes execution first without waiting for the other ones to complete execution. However, there needs to be a careful balance maintained as over-cloning requires extra resources and could lead to contention. On the other hand, under-cloning could lead to slower task execution and no effective improvement. The authors designed and experimented with short workloads with a small number of jobs. They identify that the cloning of a small number of jobs that have short execution times improves reliability without using too much additional resources. Dolly introduces a budgeted cloning strategy to only give an excess of 5\% resource consumption for a total of up to 46\% improvement in average job response time.

The work in~\cite{ananthanarayanan2014grass} proposes a strategy called Greedy and Resource Aware Speculative Scheduling (\textit{GRASS}). GRASS uses a similar strategy to Dolly, of spawning multiple clones of slow tasks but also uses greedy speculation to approximate which tasks need to be cloned, and dedicate speculation resources to improve the average deadline-bound job response time by up to 47\% and error-bound jobs by up to 38\%. The work in~\cite{zaharia2008improving} explores the MapReduce framework to investigate the occurrence of straggler tasks and optimizes its performance in a heterogeneous cloud environment. Further, the work in~\cite{wang2015using} proposes the Longest Approximate Time to End (\textit{LATE}) scheduling algorithm, which uses heuristics to search for the optimum task scheduling policy with latency and cost estimates. They also estimate the response times of all tasks of a job and assume that the one with the longest time is a straggler and execute a copy on a powerful host to reduce overall job response time. However, these works~\cite{ananthanarayanan2014grass, zaharia2008improving, wang2015using} do not adapt to dynamic environments. 

The work in~\cite{yadwadkar2014wrangler} proposes a proactive straggler management approach called \textit{Wrangler}. The underpinning predictive model uses a statistical learning technique on cluster utilization counter-data. To overcome modeling errors and maintain high reliability, Wrangler computes confidence bounds on the predictions and exploits them in the straggler management process. Specifically, Wrangler relies on a Ganglia based node monitoring to delay the execution of tasks on nodes that have straggler confidence above a threshold value. Experiments on a Hadoop-based EC2 cluster show that Wrangler is able to reduce response times by as much as 61\%, with 55\% less resources when compared to other speculative cloning based strategies. However, we show in our experiments that in certain load regimes, e.g., with low resource utilisations or with highly volatile workloads, Wrangler suffers from lower accuracy.

\textbf{Straggler Prediction and Mitigation.}
The work in~\cite{bitar2020stochastic} presents a Stochastic Gradient Coding (SGC) based approach which uses approximate gradient coding to reduce the occurrence of straggler tasks. They utilize a pair-wise balanced scheme to determine the jobs to run as a clone or redundant tasks. The SGC algorithm runs in a distributed fashion, sharing a datapoint with multiple hosts to compute independent gradients on the data which is aggregated by the master. This approach prevents the straggler analysis itself from becoming slow and hence is appropriate for volatile environments. However, in large-scale setups, monitoring data across all host machines is inefficient and can create network bandwidth contention, negatively impacting job response times.
The work in~\cite{badita2020optimal} proposes a task replication approach for job scheduling to minimize the effect of the Long-Tail problem. The authors analyze the impact of this approach in a heterogeneous platform. Their algorithm predicts the mean service times for single and multi-fork scenarios and chooses the optimal forking level. This allows their model to run multiple instances in datacenters with powerful computational resources. However, the approach can handle only a single job system with the same workload characteristics and fails in the presence of diverse workloads as pointed by~\cite{badita2020optimal}.

\section{System Model}
\label{sec:model}

We now describe the system model, which predicts the number of straggler tasks to avoid the Long Tail problem. The prediction problem requires a model to know beforehand which tasks, or at least what number of tasks may adversely impact the performance of the system. This depends on not only the types of job being executed on the CDC, but also the characteristics of the physical machines. We first discuss a Pareto distribution based model that is able to predict the number of straggler tasks based on user specifications and hyper-parameters. Later, we describe another deep learning (DL) based approach that generates these hyper-parameters of the Pareto distribution based on the characteristics of the jobs and physical cloud machines. 

\begin{figure}[]
    \centering
    \includegraphics[width=\columnwidth]{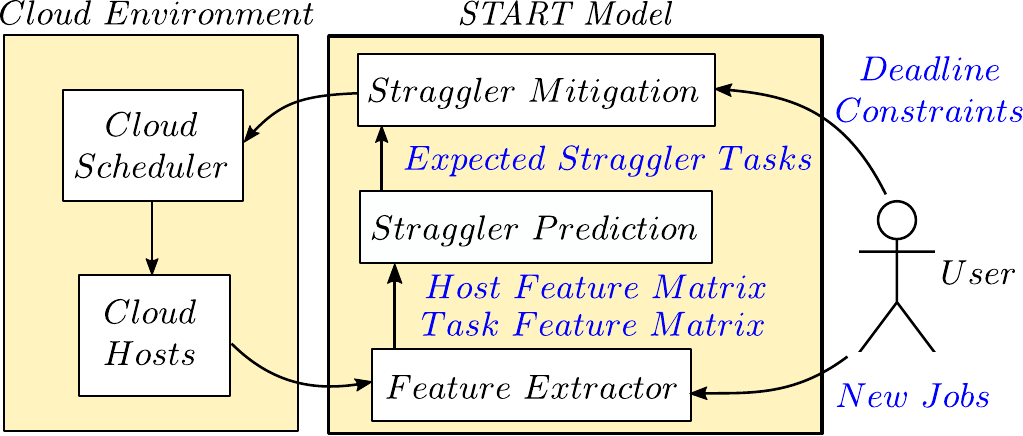}
    \caption{START System Architecture}
    \label{fig:system}
\end{figure}

A summary of our system model components and interaction is shown in Figure \ref{fig:system}. Here, the \textit{Cloud Environment} consists of a cloud scheduler and host machines. The scheduler allocates tasks onto the hosts, which are then executed and utilization metrics are captured by the resource monitoring service of the cloud environment. The utilization metrics of hosts and active tasks are then used to develop feature vectors by the \textit{Feature Extractor}. The user also provides new jobs for which the feature vectors are instantiated as $0$. The host and task feature vectors are then combined to form matrices that are then forwarded to a \textit{Straggler Prediction} module. The expected tasks flagged as stragglers by the prediction module are then mitigated using a task speculation or a re-run strategy as we describe later. 

\blue{We consider a bag-of-tasks job model where a bounded timeline is divided into equal sized scheduling intervals. At the start of each interval, the model receives a set of independent jobs. SLA deadlines are defined for each job at the time it is sent to the model. Each job consists of $q$ dependent or independent tasks, where $0< q \leq q'$.} We now describe the modeling of the response times of tasks using the Pareto distribution.

\subsection{Pareto Distribution Model}
\label{sec:pareto}

As observed in prior work such as~\cite{gill2020tails, xu2016optimization, wang2015using}, the task execution times in a cloud computing environment can be assumed to follow a Pareto Distribution for which the Cumulative Distribution Function (CDF) is
\begin{equation}
\label{eq:pareto}
\begin{aligned}
F_X(x) &= 
\begin{cases}
1-(\frac{x}{\beta})^{-\alpha} & x \geq \beta \\
0 & x < \beta ,\\
\end{cases} \\
\end{aligned}
\end{equation}
\noindent
where $\beta$ is the least time taken among tasks, and $\alpha$ is the tail index parameter ($\alpha, \beta > 0$). $X_1, X_2, \ldots, X_q$ are the times taken by $q$ tasks of a particular job running on the Cloud Environment. The Log-Likelihood Estimate~\cite{mahmoud2013estimation} is then
\begin{equation}
\label{eq:ll}
\log(L(X_1, \ldots, X_q)) = q \log(\alpha) + q\, \alpha \log(\beta) - (\alpha + 1) \sum_{i=1}^q \log(X_i),
\end{equation}
\blue{where $L$ is the likelihood function for the random variables $X_1, \ldots, X_q$.}

As $\alpha > 0$, to maximize the log likelihood, $\beta$ is obtained as the largest possible value such that $X_i > \beta,\ \forall\ i$. Thus, $\beta = \min_i(X_i)$. For $\alpha$, if we set a partial derivative of the likelihood with respect to $\alpha$ as 0, we get
\begin{equation}
\label{eq:alpha}
\begin{aligned}
\alpha &= \frac{q}{\sum_{i=1}^q \log (X_i) - q \log(\beta)}.
\end{aligned}
\end{equation}

For a given job execution, the task execution times determine the ($\alpha, \beta$) parameters of the assumed distribution. Thus, at the time of training, we run multiple jobs and fit the parameters using Equation \ref{eq:alpha}. These parameters are then used to predict the number of straggler tasks based on a straggler parameter $\mathcal{K}$, by calculating the number of tasks which in expectation could have completion times greater than $\mathcal{K}$. \blue{Thus, for $\alpha > 1$ (for a well defined mean of the distribution) and $q$ tasks, $q \cdot (1-F_X(\mathcal{K}))$ gives us the expected number of straggler tasks, where $F_X$ is the cumulative distribution function. For mathematical simplicity, we keep the straggler parameter as a multiple of the mean execution time, given as $\mathcal{K} = {k \alpha \beta} / {(\alpha - 1)}$.} This gives the expected number of straggler tasks ($E_S$),
\begin{equation}
\label{eq:straggler}
    E_S = q \left(\frac{\mathcal{K}}{\beta}\right)^{- \alpha}
\end{equation}

\begin{figure}[!t]
    \centering
    \includegraphics[width=0.9\columnwidth]{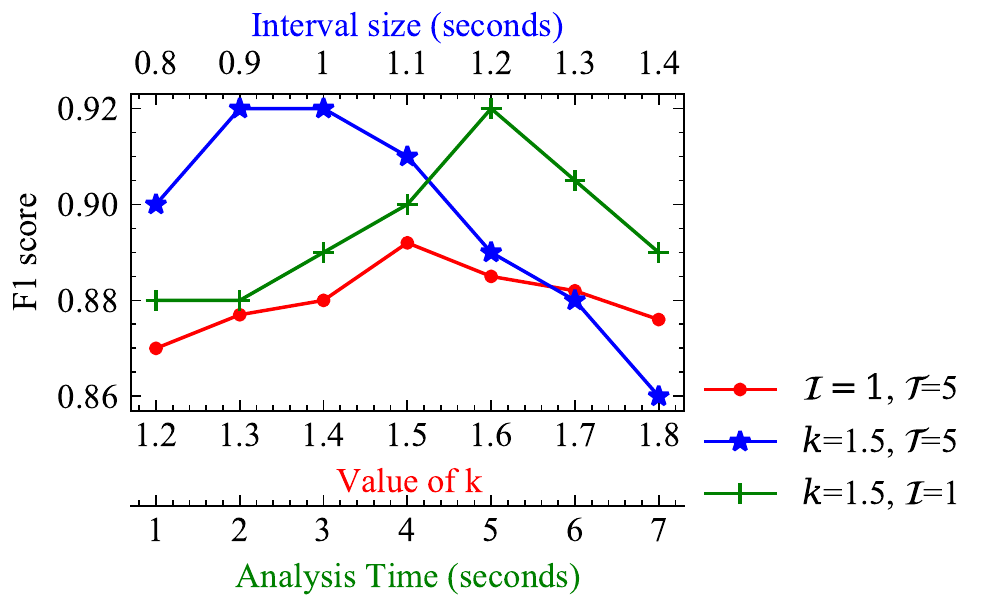}
    \caption{Empirical results for different hyper-parameter values comparing F1 scores of straggler classification on test data. $k, \mathcal{I}$ and $\mathcal{T}$ are defined in Sections~\ref{sec:pareto} and \ref{sec:encoder}. \blue{F1 score is defined as per Eq.~\ref{eq:f1}.}}
    \label{fig:emp}
\end{figure}

Empirically\footnote{As given in Figure \ref{fig:emp}, based on the method described in~\cite{yadwadkar2014wrangler} and a dataset extracted from traces on a desktop system with 64-bit Ubuntu 18.04 operating system, which is equipped with the Intel® Core™ i7-10700K processor (No. of Cores = 8, Processor Base frequency = 3.80 GHz and turbo frequency = 5.10 GHz), 64 GB of RAM, and 1 TB NVMe storage.  We have used Hadoop MapReduce for manage and execute word count application. }, we find that $k = 1.5$ strikes a good balance between the cases and hence this value is used in the experiments, but can be changed as per user requirements. \blue{Figure~\ref{fig:emp} demonstrates results corresponding to simple grid search on the three parameters $k$, $\mathcal{I}$ and $\mathcal{T}$. The latter two parameters are defined in Section~\ref{sec:encoder}. For $k = 1.5$, the prediction performance (F1 score) is the highest. For each task in the system, we check whether the predicted class is true or not, \textit{i.e.}, if the completion time of the task is $> \mathcal{K}$. The number of correct class labels is denoted as $tp$ and incorrect ones as $fp$, then the F1 score is defined as 
\begin{equation}
\label{eq:f1}
    \frac{tp}{tp + \tfrac{1}{2}(fp+tp)}.
\end{equation}
For $k<1.5$ the model has high false negatives, whereas for $k>1.5$, the model has high false positives.}

\subsection{Encoder Network}
\label{sec:encoder}
The previous subsection shows how the Pareto distribution can be used to determine the expected number of straggler tasks in a job. However, the parameters ($\alpha, \beta$) are not known beforehand for a job. As motivated in Section~\ref{sec:intro}, to predict these parameters, we use an encoder network that analyzes the tasks and the workloads at different machines in the CDC for a finite amount of time. 

\begin{figure}[!t]
    \centering
    \subfigure[$M_H$]{
    \includegraphics[width=.45\columnwidth]{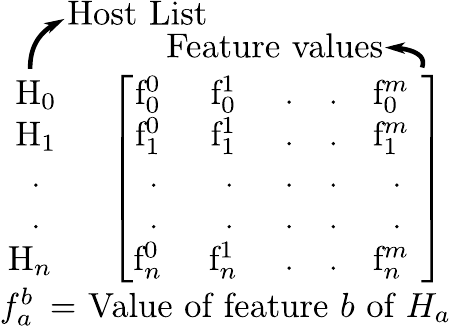}
    \label{fig:host-matrix}
    }
    \subfigure[$M_T$]{
    \includegraphics[width=.45\columnwidth]{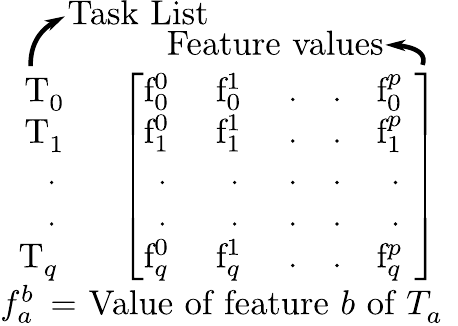}
    \label{fig:new-vm-matrix}
    }
    \caption{Matrix Representation of Model Inputs}
    \label{fig:input}
\end{figure}

\begin{figure}[t]
    \centering \setlength{\belowcaptionskip}{-10pt}
    \includegraphics[width=\columnwidth]{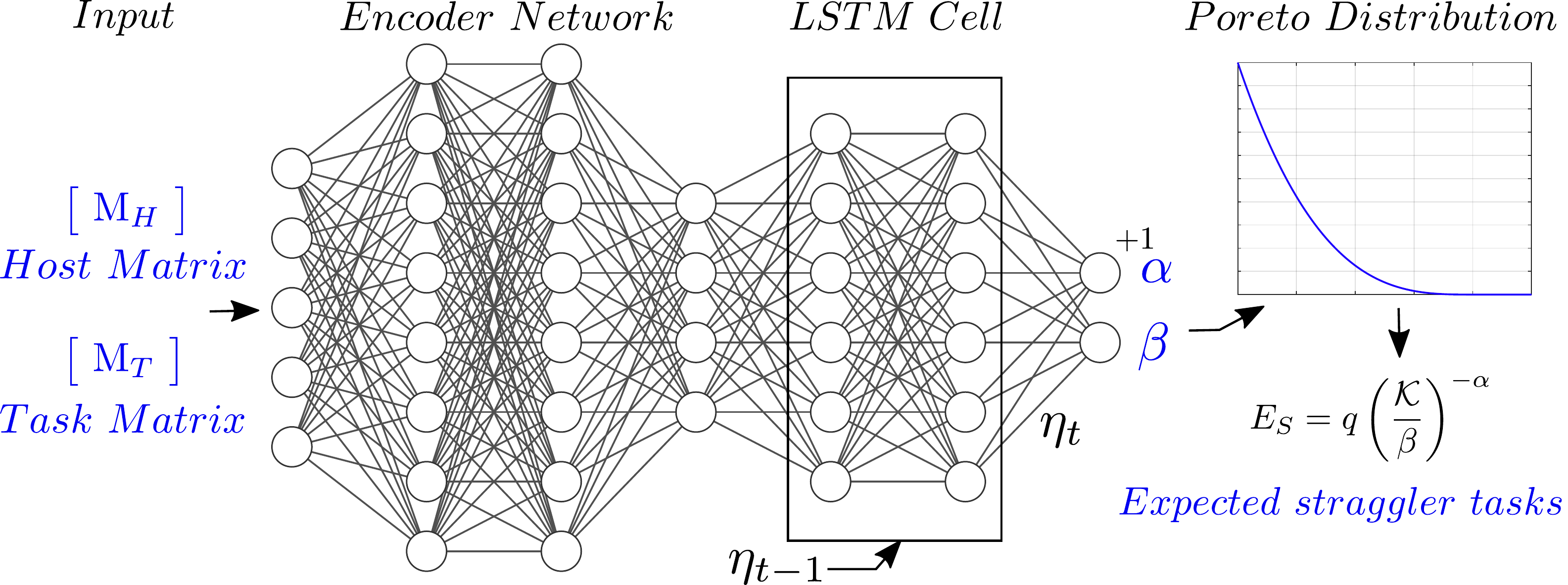}
    \caption{Straggler prediction model}
    \label{fig:model}
\end{figure}

We first identify a job $j$ as a set of tasks \blue{\{$T_i$\}$_{i=1}^{q}$, where $q<q'$ if less than $q'$ tasks then rest $q'-q$ rows are 0.} For each task $T_a$, $p$ feature values are used to form a feature vector. Similarly, for each host out of $n$ hosts \{$H_i$\}$_{i=1}^n$, $m$ feature values are used. The features used for hosts include utilization and capacity of CPU, RAM, Disk and network bandwidth. The feature vector also includes the cost, power characteristics, and the number of tasks to which this host is allocated. The features used for tasks include CPU, RAM, Disk and bandwidth requirements and the host assigned in the previous interval. These were used to characterize the system state for deep learning models as is common in prior art \cite{tuli2020dynamic, zhu2019novel, aktacs2019straggler}. These feature vectors of hosts ($M_H$) and tasks ($M_T$), as shown in Figure \ref{fig:input}, are then used to predict the Pareto parameter values. The neural network model and the working of the system is shown in Figure \ref{fig:model}. The input matrices are first passed through an encoder network, the output of which is sent to a Long Short Term Memory (LSTM) network~\cite{gers1999learning}. To prevent the LSTM model from diverging, we take an exponential moving average of each matrix using a $0.8$ weight to the latest resource matrix (as in~\cite{lin2016hybrid}). For time-series prediction, multiple machine learning models could be used, including Echo State Networks (ESN) or LSTMs~\cite{shuja2020applying}. However, as ESNs control the degree of delays using a manually chosen constant (leaking rate), this typically lowers the generalization ability when applied to different load traces~\cite{song2018host}. Hence, we use LSTMs to develop our parameter estimation model.

The Encoder network is a 4 layer fully-connected network with the following details (adapted from prior art~\cite{tuli2019healthfog, tuli2020dynamic, zhu2019novel}):
\begin{itemize}
    \item Input layer of size $|M_H|+|M_T|$. The non-linearity used here is $\textsf{softplus}$\footnote{The definitions of these activation functions can be seen at the PyTorch web-page: \texttt{https://pytorch.org/docs/stable/nn.html}} as in \cite{tuli2020dynamic}. The matrices are flattened, concatenated and given as an input to the encoder network.
    \item Fully connect layer of size $128$ with $\textsf{softplus}$ activation.
    \item Fully connect layer of size $128$ with $\textsf{softplus}$ activation.
    \item Fully connect layer of size $32$ with $\textsf{softplus}$ activation.
\end{itemize}

\blue{We run inference using a neural network model for each job. Specifically, for each job $j$, we provide the model with the inputs $M_H$ for host characteristics and $M_T$ for all running tasks in $j$. For each job, we generate $\alpha, \beta$ parameters of the Pareto distribution to evaluate the number of straggler tasks.} The LSTM network has 2 layers with size 32 nodes. The predicted output of the LSTM network becomes an input for a fully connected layer with 2 nodes, which outputs the ($\alpha, \beta$) values after a Rectified Non-linear Unit (ReLU) so that these values are positive (with addition of 1 to $\alpha$ so that the mean of the distribution is defined). This is sent to the LSTM Network. To implement the proposed approach, we use PyTorch Autograd package~\cite{paszke2017automatic} to run the back-propagation procedure for network training. \blue{We keep sending the input matrices for a finite time of $\mathcal{T}$, periodically after every $\mathcal{I}$ seconds. The LSTM cell takes in two inputs, the hidden state of the previous interval and the output of the encoder network.} Considering the output of the previous iteration, \emph{i.e.}, the hidden state $\eta_{t-1}$ and the output of the encoder network $\lambda$, the output for the current interval becomes $\eta_t = LSTM(\eta_{t-1}, \lambda)$ (see Figure~\ref{fig:model}). Here, $\eta_0 = 0$ and $t \in \{0, \mathcal{I}, 2\cdot\mathcal{I}, \ldots, \mathcal{T}\}$. \blue{Using grid-search, for the experiments we set $\mathcal{I} = 1$ and $\mathcal{T} = 5$, which empirically gives the best results\footnotemark[1]. }

\begin{table}[]
    \centering
    \caption{\blue{Notation}}
    \begin{tabular}{@{}ll@{}}
        \toprule
        Symbol & Meaning \\
        \midrule
        $q$ & Maximum number of tasks in a job \\ 
        $\alpha, \beta$ & Parameters of the Pareto distribution \\ 
        $\mathcal{K}$ & Straggler parameter in START \\ 
        $E_S$ & Expected number of straggler tasks \\ 
        $\mathcal{I}$ & Time-period of START inference in seconds \\
        $\mathcal{T}$ & Time-duration of START inference in seconds \\
        $n$ & Number of hosts \\ 
        \bottomrule
    \end{tabular}
    \label{tab:symbols}
\end{table}

The output of LSTM network gives us the parameters for the Pareto distribution, which are then used to find expected straggler tasks ($E_S$). This constitutes the \textit{Straggler Prediction} module in Figure~\ref{fig:system}. The objective of the model training is to predict the appropriate distribution parameters using the utilization metrics and use this distribution to calculate the expected number of straggler tasks as described in Section~\ref{sec:pareto}. $E_S$ determines the number of tasks to mitigate using rerun/speculation-based methods, as explained in the next subsection. Out of the $q$ tasks, first the parameters ($\alpha, \beta$) are calculated after $\mathcal{T}$ time-steps and then $\floor{E_S}$ tasks are mitigated. This ensures that if $E_S$ is very small ($<1$), we do not mitigate any tasks, saving computational resources. Hence, after execution of $q - \floor{E_S}$ tasks, we apply mitigation techniques on the remaining tasks to prevent delays in result generation. Compared to other methods, our model nearly eliminates the detection time and hence is able to provide a faster response to users (as shown in Section~\ref{sec:exp}). \blue{The main symbols and their meanings are summarized in Table~\ref{tab:symbols}.}

\subsection{Speculation and Task Rerun}

\begin{algorithm}[!t]
\caption{Straggler Prediction and Mitigation Algorithm}
\begin{algorithmic} [1]
\Statex \textbf{Inputs:}
\State $J \leftarrow$ Set of all jobs being executed currently $[j_1, j_2, ... , j_r]$
\State $T^n_m \leftarrow$ Set of tasks of job $j_n$ where $m \in \{1, 2, 3 ... q\}$
\State $M_{time} \leftarrow$ Max allocated time to release the resource. 
\Statex \textbf{Variables:}
\State $J_n \leftarrow$ Set of normal jobs $\subseteq J$ without straggler tasks
\State $J_s \leftarrow$ Set of jobs $\subseteq J$ with $>0$ straggler tasks
\Statex \textbf{Procedure} \textsc{PredictStraggler}(job)
\State \hspace{\algorithmicindent}\textbf{for} time t from 0 to $\mathcal{T}$ with step $\mathcal{I}$
\State \hspace{\algorithmicindent}\hspace{\algorithmicindent}$q \leftarrow$ Number of tasks in input job
\State \hspace{\algorithmicindent}\hspace{\algorithmicindent}Extract feature vectors of host machines as $M_H$
\State \hspace{\algorithmicindent}\hspace{\algorithmicindent}Extract feature vectors of tasks of input job as $M_T$
\State  \hspace{\algorithmicindent}Predict $(\alpha, \beta)$ using the Neural network 
\State  \hspace{\algorithmicindent}Find $E_S$ as $ q \left(\frac{\mathcal{K}}{\beta}\right)^{- \alpha}$
\State  \hspace{\algorithmicindent}Run job till completion of $q - \floor{E_S}$ tasks
\State  \hspace{\algorithmicindent}\textbf{return} incomplete tasks
\State \textbf{Procedure} \textsc{Speculation}(task list)
\State \hspace{\algorithmicindent}\textbf{for} {task $t$ in task list}
\State \hspace{\algorithmicindent}\hspace{\algorithmicindent}Run a copy of $t$ on a different node
\State \textbf{Procedure} \textsc{ReRunStragglerTask}(task list)
\State \hspace{\algorithmicindent}\textbf{for} {task $t$ in task list}
\State \hspace{\algorithmicindent}\hspace{\algorithmicindent}Run the same task $t$ on different node
\State \textbf{Begin}
\State \textbf{for} {job $j_i$ in $J$}
\State \hspace{\algorithmicindent}$stragglerTasks \leftarrow \textsc{PredictStraggler}(j_i)$
\State \hspace{\algorithmicindent}\textbf{if} {$stragglerTasks$ is empty}
\State \hspace{\algorithmicindent}\hspace{\algorithmicindent}add $j_i$ to $J_n$
\State \hspace{\algorithmicindent}\hspace{\algorithmicindent}\textbf{continue}
\State \hspace{\algorithmicindent}\textbf{else}
\State \hspace{\algorithmicindent}\hspace{\algorithmicindent}add $j_i$ to $J_s$
\State \hspace{\algorithmicindent}\hspace{\algorithmicindent}Wait for specific time ($M_{time}$), if $j_i$ does not respond then generate alert for action 
\State \hspace{\algorithmicindent}\textbf{if} {$j_i$ is deadline oriented}
\State \hspace{\algorithmicindent}\hspace{\algorithmicindent}$\textsc{Speculation}(stragglerTasks)$
\State \hspace{\algorithmicindent}\textbf{else}
\State \hspace{\algorithmicindent}\hspace{\algorithmicindent}$\textsc{ReRunStragglerTask}(stragglerTasks)$
\end{algorithmic}
\label{alg:scheduling}
\end{algorithm}

To mitigate the Long Tail problem, we use the following two strategies (as in prior work~\cite{gill2020tails, badita2020optimal}) for the straggler tasks detected by our prediction model.
\begin{enumerate}
    \item \textbf{Speculation:} We run a copy of the straggler task on a separate node and use the results we get first. This is crucial for deadline driven tasks that need results as soon as possible. Thus, this method gives us the least response time at the cost of running multiple nodes.
    \item \textbf{Re-Run Task:} We stop execution of the straggler task on the respective node and run a new instance of the same task in a new node. This method is suitable for tasks that are not deadline critical as it runs only one copy of the task at a time which reduces energy consumption and prevents congestion.
\end{enumerate}

\begin{figure}[!t]
    \centering \setlength{\belowcaptionskip}{-10pt}
    \includegraphics[width=0.62\columnwidth]{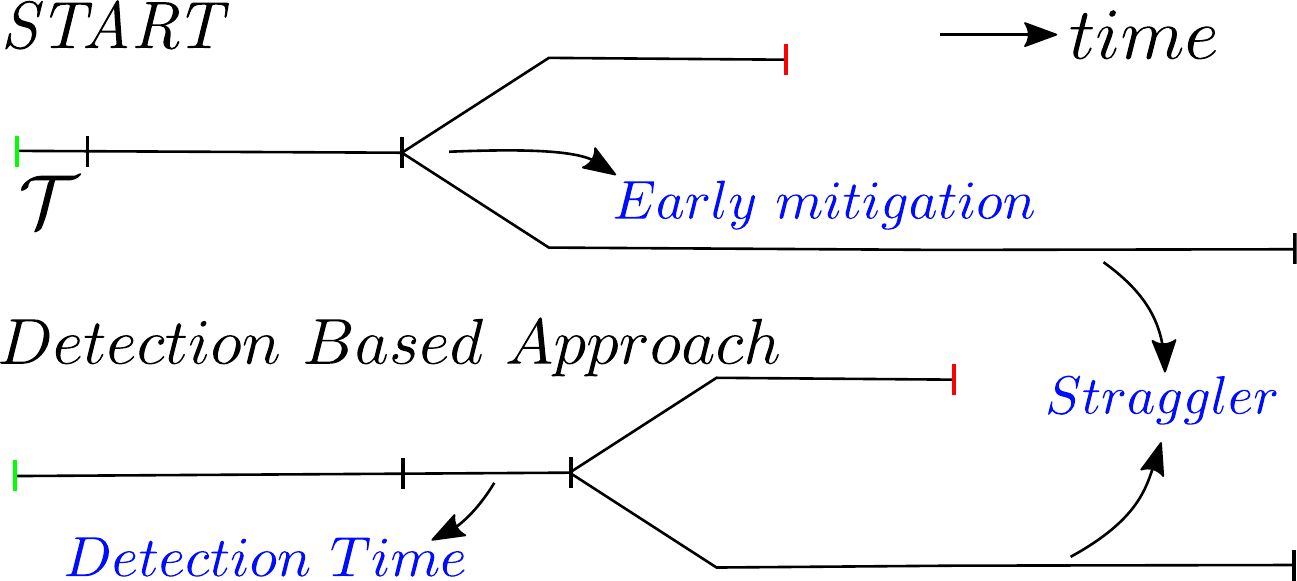}
    \caption{Comparison of START with detection based approaches.}
    \label{fig:comparison}
\end{figure}

The choice of the separate or new node is performed by the underlying scheduling scheme (further details in Section~\ref{sec:setup}). We do not consider task cloning as it has significant overheads in large-scale environments~\cite{garraghan2016straggler}. In both approaches mentioned above, we select the new node that has the lowest moving average of the number of straggler tasks for the current time-step. Algorithm \ref{alg:scheduling} describes in detail the complete approach of straggler prediction and mitigation and is run periodically to eliminate the long tail problem. As shown, START first determines the host and task feature matrices for every job (lines 8 and 9), which are then analyzed for $\mathcal{T}$ time-steps to predict the number of straggler tasks (line 13). For each job which has $\floor{E_S} > 0$, mitigation techniques are run for remaining tasks when only $\floor{E_S}$ of them are left (lines 30 and 32). Figure \ref{fig:comparison} shows how START is able to provide much lower response times compared to existing detection based algorithms by nearly eliminating the detection time as it predicts early-on the number of tasks that are highly likely to be stragglers. This constitutes the \textit{Straggler Mitigation} module in Figure~\ref{fig:system}.

\section{Evaluation Setup}
\label{sec:setup}
\subsection{Evaluation Metrics}
\label{sec:metrics}
We use common evaluation metrics~\cite{ananthanarayanan2014grass, gill2020tails, bitar2020stochastic}. We assume there are $n$ \blue{host} and $\blue{q}$ jobs currently in the system.

\noindent
\textit{1) Energy Consumption:} The cumulative energy consumed for a given time is given by
\begin{equation}
    E = E_{CPU} + E_{Disk} + E_{Memory} + E_{Network} + E_{Misc},
    \label{eq:energy}
\end{equation}
\noindent
where $E_{CPU}$ is the total energy consumed by all the processors, which includes dynamic energy as $CV^2f$, short-circuit energy, leakage energy, and idle energy consumption~\cite{gill2019holistic}. $E_{Disk}$ is the energy consumed for all read/write operations plus the idle energy consumed by all the disks. $E_{Memory}$ is the energy consumed by all memories (RAM and Cache) in the computational nodes. $E_{Network}$ is sum of energies consumed by network devices which include routers, gateways, LAN cards and switches. $E_{Misc}$ is energy consumed by other components like motherboard and port connectors. However, in simulation it is difficult to find out each energy component separately, so we calculate maximum and minimum energy consumption ($E_{max}, E_{min}$) by hardware profiling as per Equation \ref{eq:energy} and using Standard Performance Evaluation Corporation (SPEC) benchmarks \texttt{https://www.spec.org/cloud\_iaas2018/results/}. We then use Equation \ref{eq:energy2} to get total energy consumption in CloudSim at time $t$.  Here, $U_k^t$ is the total \blue{host} resource utilization (sum of all workloads) of \blue{host} $k$. This is a common practice~\cite{calheiros2011cloudsim}. Thus,
\begin{equation}
    E_{total}^t = \sum_{k=1}^n U_k^t \cdot (E_{max} - E_{min}) + E_{min}.
    \label{eq:energy2}
\end{equation}

\noindent
\textit{2) Execution Time:} The average execution time is
\begin{equation}
    T_{avg}^{exec} = \frac{1}{\blue{q}} \sum_{i=1}^{\blue{q}} (T_i^C - T_i^S) +  \sum_{i=1}^{\blue{q}} R_i.
    \label{eq:exec}
\end{equation}

This is the total time taken to successfully execute an application, on average, for all tasks. Here $T_i^C$, $T_i^S$ and $R_i$ are the completion, submission and restart time of task $i$.

\noindent
\textit{3) Resource Contention:} Resource contention occurs when one workload shares the same resource during the execution [20]. This may be due to unavailability of the required number of resources, or because there are a large number of workloads with urgent deadlines. Resource contention is quantified as
\begin{equation}
    Con_{total}^{resource} = \sum_{k=1}^n \sum_{i=1}^{\blue{q}_k} Req_{i,k}^{resource} \cdot \mathds{1}(resource_i\ overloaded),
    \label{eq:contention}
\end{equation}
where $\blue{q}_k$ is the number of tasks being executed at resource $k$ and $Req_{i,k}^{resource}$ is the resource requirement of $i^{th}$ task at node $k$. Also, $\mathds{1()}$ denotes the indicator function. 

\noindent
\textit{4) Memory Utilization:} The memory utilization of \blue{host} $k$ in percentage terms is
\begin{equation}
    U_{k}^{memory} = \frac{P^{total}_k - (F_k + B_k + C_k)}{P^{total}_k} \times 100,
    \label{eq:memory}
\end{equation}
where $P_k^{total}, F_k, B_k, C_k$ are the total physical, free, buffer and cache memory respectively.

\noindent
\textit{5) Disk Utilization:} The disk utilization of \blue{host} $k$ in percentage terms is
\begin{equation}
    U_{k}^{disk} = \frac{Total\, Used}{Total\, HD\, Size} \times 100.
    \label{eq:disk}
\end{equation}

\noindent
\textit{6) Network Utilization:} The network utilization of \blue{host} $k$ in percentage terms is
\begin{equation}
    U_{k}^{network} = \frac{Bits_{total}^{rx} + Bits_{total}^{tx}}{BW_k \times S_I} \times 100,
    \label{eq:network}
\end{equation}
where $Bits_{total}^{rx}$ and $Bits_{total}^{tx}$ are the total bits received and transmitted in an interval. $BW_k$ is the bandwidth of \blue{host} $k$ and $S_I$ is the size of the interval.

\noindent
\textit{7) SLA Violation Rate:} For $\blue{q}$ tasks we have $\blue{q}$ SLAs. Each SLA has a weight ($i^{th}$ SLA having weight $w_i$). The total SLA violation rate is
\begin{equation}
    SLA_{total}^{violation} = \frac{1}{\blue{q}} \sum_{i=1}^{\blue{q}} w_i \cdot \mathds{1}(SLA_i\ \text{is violated}).
    \label{eq:sla}
\end{equation}
We also use other metrics including \textit{Resource contention}, \textit{CPU utilization} and \textit{Completion times} as defined in \cite{gill2019radar}. 

\blue{As per prior work~\cite{gill2020tails}, the metric for comparing prediction accuracy is the Mean Average Percentage Error (MAPE) which is defined as the mean percentage error of the predicted value (number of straggler tasks for each job) from the actual value and given by Equation \ref{mape}. To obtain the actual value, we only perform straggler prediction and compare MAPE of START, IGRU-SD and RPPS \cite{fang2012rpps} as other baselines do not perform straggler prediction. We use this to calculate the number of straggler tasks using maximum-likelihood estimation (see Equation~\ref{eq:straggler}). Thus,
\begin{equation}
\label{mape}
    MAPE = \frac{100\%}{n} \sum_{t=1}^{n} \left| \frac{y_t - y_t'}{y_t} \right|,
\end{equation}
where $y_t$ and $y_t'$ are the actual and predicted number of straggler tasks and $n$ is the number of scheduling intervals for the complete simulation.}

\begin{table*}[!t]
    \centering
    \caption{Configuration Details of simulated Physical machines}
    \begin{tabular}{@{}lcccc@{}}
    \toprule 
    \textbf{CPU} & \textbf{RAM and Storage} & \textbf{Core count} & \textbf{Operating System} & \textbf{Number of Virtual Nodes}\tabularnewline
    \midrule
    Intel Core 2 Duo - 2.4 GHz & 6 GB RAM and 320 GB HDD & 2 & Windows & 12\tabularnewline
    Intel Core i5-2310- 2.9GHz & 4 GB RAM and 160 GB HDD & 4 & Linux & 6 \tabularnewline
    Intel XEON E 52407-2.2 GHz & 2 GB RAM and 160 GB HDD & 4 & Linux & 2\tabularnewline
    \bottomrule 
    \end{tabular}
    \label{tab:setup}
\end{table*}

\subsection{Workload Model}

Our evaluation uses CloudSim toolkit and real-time workload traces are derived from PlanetLab systems~\cite{park2006comon}. This dataset contains traces of CPU, RAM, disk, and network bandwidth requirements from over 1000 PlanetLab tasks collected during 10 random days. These traces are collected using a scheduling interval size of 300 seconds. The virtual machines are located at more than 500 places across the globe. The data was collected on 2880 intervals each, thus each trace was of this size~\cite{kim2011understanding}\footnote{The traces from the PlanetLab systems can be downloaded from \texttt{https://www.planet-lab.org/planetlablogs}.}.
In this dataset, 50\% of the traces are deadline driven and 50\% are not. We get similar results on other distributions. A collection of 2 to 10 tasks is defined as a job. We use data for 800 tasks as our training set and 100 tasks' data as the test set. As in prior work~\cite{tuli2020dynamic}, a Poisson Distribution $Poisson(\lambda)$, with $\lambda = 1.2$ jobs, is selected for the number of jobs to be created periodically. This is because all the workloads/tasks of different jobs are independent of each other. The requests submitted by users are considered as cloudlets, which have three specific requirements (CPU, memory and task length).

\subsection{CloudSim Simulation Environment}

We evaluate the performance of START using a simulated cloud environment. We implement our straggler detection and mitigation technique by introducing the different kinds of faults using an event-driven module. The neural network and back-propagation through time code were implemented using PyTorch library in Python. As in prior work~\cite{nita2014fim}, we have used a Weibull Distribution to model failure characteristics. The failure distribution is given by
\begin{equation}
    f(x; k, \lambda) = \frac{k}{\lambda}\left(\frac{x}{\lambda}\right)^{k-1},
\end{equation}
where $x$ is the time-to-failure. We assign the parameters $k = 1.5, \lambda = 2$ as in~\cite{nita2014fim, zheng2018hound}. The introduced fault types are (1) host \blue{faults} (memory \blue{faults} and \blue{faults} in the processing elements), (2) Cloudlet \blue{faults} (due to network faults) and (3) VM creation \blue{faults}. \blue{We consider task faults where the underpinning applications need to rerun due to task breakdown. For host failure, all tasks running in that host need to restart. We consider only ephemeral host faults, \textit{i.e.}, our hosts are offline for a short duration of time (up to 4 intervals in our experiments) instead of being permanently down. Other faults considered in the system include unavailability of memory space, disk page faults and network packet drops that increase the response time of running tasks.} Every change in the states of VMs and hosts should be realized by the cloud datacenter through the cloud broker. Further, the broker uses a cloudlet specification to request the creation of VM and scheduling of cloudlets. We have designed a Fault Injection Module to create a fault injector thread by simulating the cloudlet \blue{faults}, host \blue{faults} and VM creation \blue{faults}. A failed node can return to service only after a downtime as defined in~\cite{nita2014fim}. 

The Fault injector thread uses a Weibull Distribution and generates events which execute commands such as \textit{“sendNow(dataCenter.getId(), FaultEventTags.HOST\_FAILURE, host);”}~\cite{nita2014fim}. The Fault Injection Module contains three entities such as \textit{FaultInjector}, \textit{FaultEvent} and \textit{FaultHandlerDatacenter}. \textit{FaultInjector} extends the SimEntity class of CloudSim and start simulation to insert fault events randomly using the Weibull Distribution. \textit{FaultEvent} extends the SimEvent class of CloudSim, which describes the type of faults such as create VM failure, cloudlet failure and host failure. \textit{FaultHandlerDatacenter} extends the Datacenter class and processes fault events sent by the \textit{FaultGenerator} and handles VM migration. In this simulation setup, four Physical Machines (PMs) characteristics (CPU, RAM, Disk and Bandwidth capabilities) are used with a various number of virtual nodes as shown in Table~\ref{tab:setup}. Since straggler tasks are particularly common in resource-constrained devices~\cite{gill2020tails}, we use devices with low core count and RAM for our experiments. The test setup is similar to prior work~\cite{gill2019radar} .

Table \ref{tab:params} details the values of the simulation parameters used in the performance evaluation, collected from the existing literature and empirical studies~\cite{gill2019holistic, li2018holistic, kouki2012sla, balis2018holistic}. We keep the parameters $\mathcal{I}$ and $\mathcal{T}$ fixed as 1 and 5 seconds respectively throughout the simulation. We dynamically change the $k$ value based on empirical results for the data up till the current interval with the initial value as $1.5$ (as described in Section~\ref{eq:pareto}).

\begin{table}[]
    \centering
    \caption{Simulation Parameters for experiments}
    \begin{tabular}{@{}lc@{}}
    \toprule 
    \textbf{Parameter } & \textbf{Value}\tabularnewline
    \midrule
    Number of VMs (n) & 400\tabularnewline
    Number of Cloudlets (Workloads) & 5000\tabularnewline
    Host Bandwidth & 1 -2 KB/S\tabularnewline
    CPU IPS (in millions) & 2000\tabularnewline
    Cloud Workload size & 10000 $\pm$ 3000 MB\tabularnewline
    Cloud Workload cost & 3 - 5 C\$ \tabularnewline
    Memory Size & 2-12 GB\tabularnewline
    Input File size & 300 $\pm$ 120 MB\tabularnewline
    Output File size & 300 $\pm$ 150 MB\tabularnewline
    Power Consumption (KW) & 108 - 273 KW\tabularnewline
    Latency of hosts  & 20-90 Seconds\tabularnewline
    Size of Cache memory  & 4 - 16 MB\tabularnewline
    CPU Power Consumption  & 130 - 240W\tabularnewline
    RAM Power Consumption & 10 - 30W\tabularnewline
    Disk Power Consumption & 3 - 110W\tabularnewline
    Network Power Consumption & 70 - 180W\tabularnewline
    Power Consumption of other Components  & 2 - 25W\tabularnewline
    \bottomrule 
    \end{tabular}
    \label{tab:params}
\end{table}

\subsection{Model Training}

To train the Encoder-LSTM network, we use the PlanetLab dataset and divide the workloads of 1000 tasks into 80\% training dataset and the rest as the test dataset. For training and test sets too, we keep the 50-50 ratio of tasks that are deadline-driven to those that are not. Further, we use a scheduler that selects tasks at random and schedules them randomly to any host using a uniform distribution. The random scheduler allows us to obtain diverse host and task characteristics for model training, which is crucial to prevent under-fitting of the neural network. The response time histogram was generated and compared against the $(\alpha, \beta)$ output of the Encoder-LSTM network. The model was trained using Mean-Square-Error Loss between the values based on the predicted distribution and the actual data. We used a learning rate of $10^{-5}$ and the Adam optimizer to train the network~\cite{kingma2014adam}.

\subsection{VM Scheduling Policy}

\blue{We use the A3C-R2N2 policy which schedules workloads using a policy gradient based reinforcement learning strategy which tries to optimize an actor-critic pair of agents~\cite{tuli2020dynamic}. This approach uses Residual Recurrent Neural Networks (R2N2) to predict the expected reward for each action (i.e scheduling decision) and tries to optimize the cumulative reward signal. The A3C-R2N2 policy has been shown to outperform other policies in terms of response time and SLA violations~\cite{tuli2020dynamic}; hence, it is our choice of scheduling method for comparing straggler mitigation techniques.}

\subsection{Baseline Algorithms}

\begin{figure*}[!t]
{
    \centering \setlength{\belowcaptionskip}{-10pt}
    \includegraphics[width=.6\textwidth]{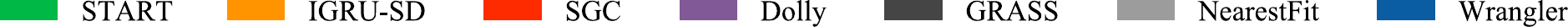}
    \\
    \subfigure[]{
    \includegraphics[width=.23\textwidth]{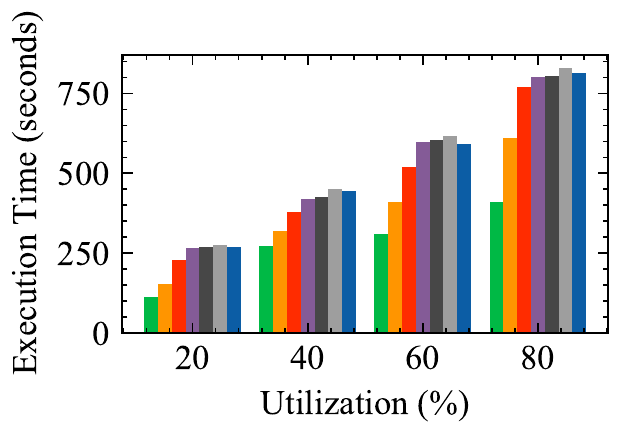}
    \label{fig:4a}
    }
    \subfigure[]{
    \includegraphics[width=.23\textwidth]{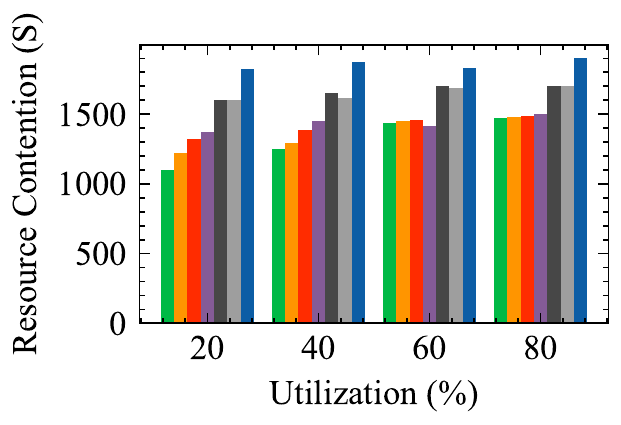}
    \label{fig:4b}
    }
    \subfigure[]{
    \includegraphics[width=.23\textwidth]{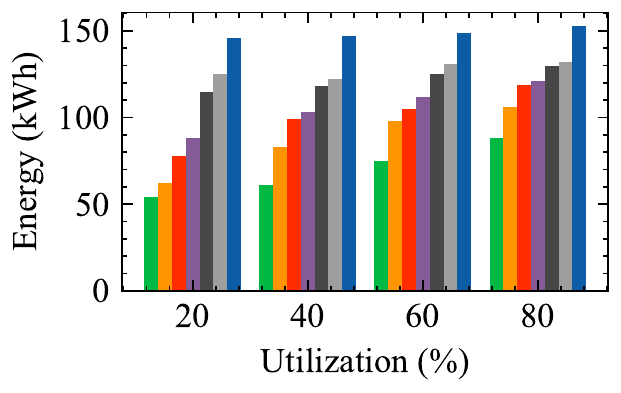}
    \label{fig:4c}
    }
    \subfigure[]{
    \includegraphics[width=.23\textwidth]{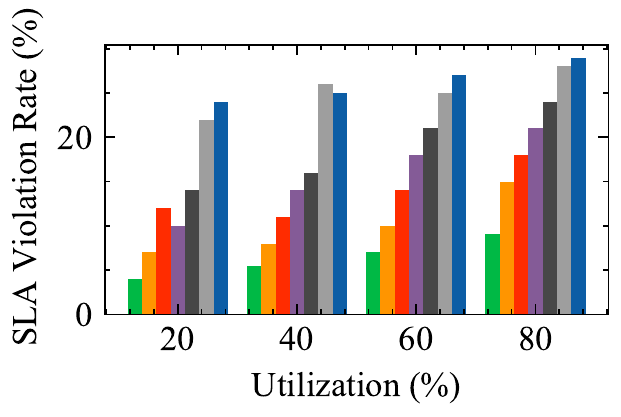}
    \label{fig:4d}
    }
    \caption{Comparison of QoS parameters with different value of CPU, disk, network and memory Utilization: a) Execution Time, b) Resource Contention, c) Energy Consumption and d) SLA Violation Rate}
    \label{fig:comparison2}
}
\end{figure*}

We have selected six baseline techniques \textit{NearestFit}, \textit{Dolly}, \textit{GRASS}, \textit{SGC}, \textit{Wrangler} and \textit{IGRU-SD} which are the most recent among prior works (see Section~\ref{sec:relwork} for details). We have chosen recent and relevant techniques from the literature to validate our technique against state-of-the-art techniques.

\begin{enumerate}[leftmargin=*]
    \item \textit{NearestFit:} uses a statistical curve fitting approach to detect stragglers. The function $a + b\cdot x^c$ is fitted with $x$ as the size of the input file for a task~\cite{coppa2015data}. However, vanilla NearestFit is not able to mitigate the detected stragglers, so we use speculation on the detected tasks.
    \item \textit{Dolly:} is a straggler mitigation technique that forks tasks into multiple clones which are executed in parallel within their specified budget. The number of clones are calculated based on the Upper-Confidence-Bound as in~\cite{ananthanarayanan2013effective} using the CPU utilization of tasks. 
    \item \textit{GRASS:} is straggler mitigation framework, which uses the concept of speculation to mitigate stragglers reactively. It is implemented using two algorithms, one for greedy speculation and the other for resource-aware scheduling.
    \item \textit{SGC:} is an approach using distributed gradient calculation to utilize a pair-wise balancing scheme for running clones of tasks.
    \item \textit{Wrangler:} is a proactive straggler mitigation technique, which uses linear modelling approach to reduce the utilization of excess resources by delaying the start of tasks predicted as straggler. 
    \item \textit{IGRU-SD:} is a GRU neural network based resource requirement prediction technique which uses detection mechanisms on the predicted future characteristics~\cite{lu2019gru}. As it only predicts straggler tasks and does not mitigate them, we use the same re-run and speculation strategy (based on deadline requirements) for fair comparison.
\end{enumerate}


\section{Performance Evaluation}
\label{sec:exp}

\subsection{Experimental Observations}

As in prior work~\cite{gill2020tails, badita2020optimal}, we used QoS parameters to evaluate the performance of START as compared to the existing techniques. We run our experiments for 24 hours, \emph{i.e.}, 288 scheduling intervals. We average over 5 runs and use diverse workload types to ensure statistical significance.

\subsubsection{Variation of Resource Utilization}

We consider 4 types of reserved utilization for CPU, disk, memory and network, where utilization is blocked intentionally (20\%, 40\%, 60\% and 80\%) to test the performance of the proposed technique. Figure~\blue{\ref{fig:comparison2}} shows the comparison of QoS parameters such as Execution Time, Energy, Resource Contention and SVR with different values of CPU, disk, network and memory utilization. 

Figure \ref{fig:4a} shows the value of execution time for different straggler management techniques with variation in the value of CPU, disk, network and memory utilization. The value of execution time increases with the increase in the value of reserved utilization, but START performs better than the existing techniques because it tracks the states of the resources dynamically for efficient decisions. The value of execution time in START is 11.47-17.4\% less than the baseline methods. 
Figure \ref{fig:4b} shows the variation of resource contention with different values of utilization. The value of resource contention increases as the value of utilization increases. The value of resource contention in START is 12.34-15.19\% less than the baseline methods. This is due to the execution time variation across various tasks and resources due to the filtered resource list obtained from the resource provisioning unit (see Section~\ref{sec:relwork}). 

Figure \ref{fig:4c} shows the energy consumption for different values of utilization and we observe that energy consumption increases with the utilization for all straggler management techniques. However, START performs better than the prior art as it avoids over or under-utilization of resources during scheduling. The value of energy consumption in START is between 18.55\% and 22.43\% less than the baseline methods. 
Figure \ref{fig:4d} shows the variation of SLA violation rate with different values of utilization and value of SLA violation rate is increasing as the value of utilization increases. The value of SLA violation rate in START is between 21.34\% and 26.77\% less than the baseline methods. This occurs because START uses admission control and a reservation mechanism for execution of workloads in advance.

\begin{figure*}[!t]
{
    \centering \setlength{\belowcaptionskip}{-12pt}
    \includegraphics[width=.6\textwidth]{graphs/legend.pdf}
    \\
    \subfigure[]{
    \includegraphics[width=.23\textwidth]{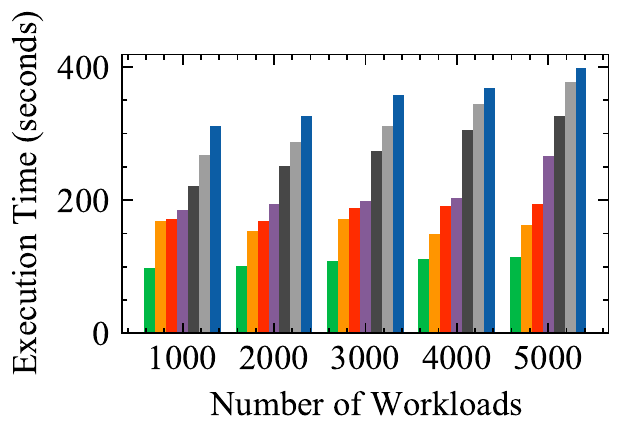}
    \label{fig:5a}
    }
    \subfigure[]{
    \includegraphics[width=.23\textwidth]{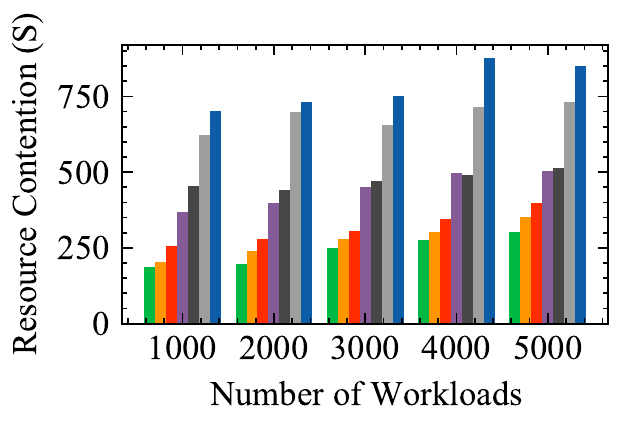}
    \label{fig:5b}
    }
    \subfigure[]{
    \includegraphics[width=.23\textwidth]{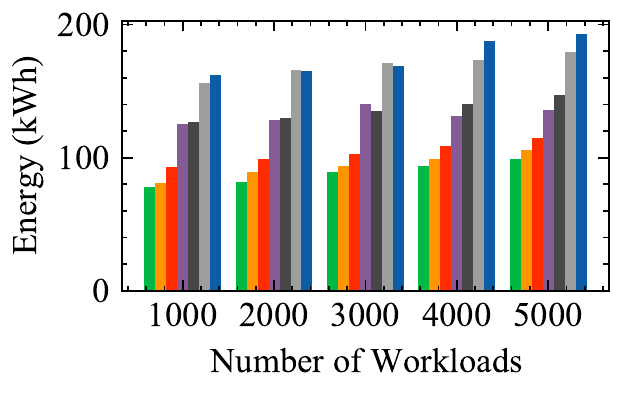}
    \label{fig:5c}
    }
    \subfigure[]{
    \includegraphics[width=.23\textwidth]{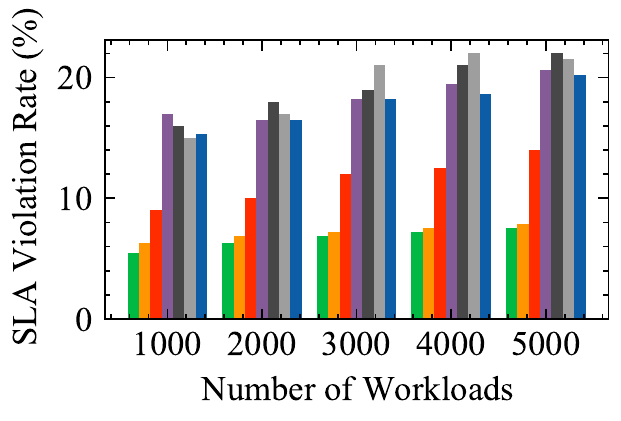}
    \label{fig:5d}
    }\\
    \subfigure[]{
    \includegraphics[width=.23\textwidth]{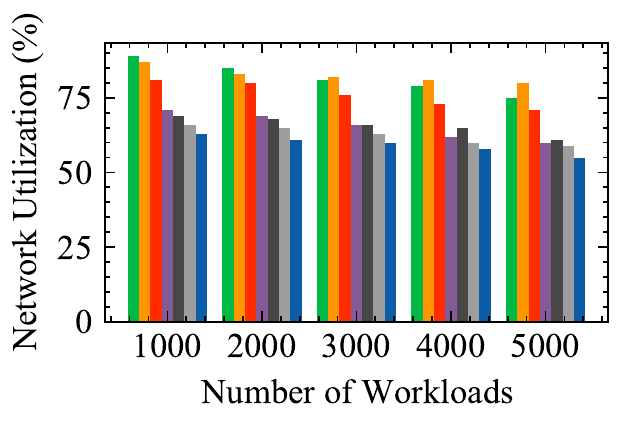}
    \label{fig:5e}
    }
    \subfigure[]{
    \includegraphics[width=.23\textwidth]{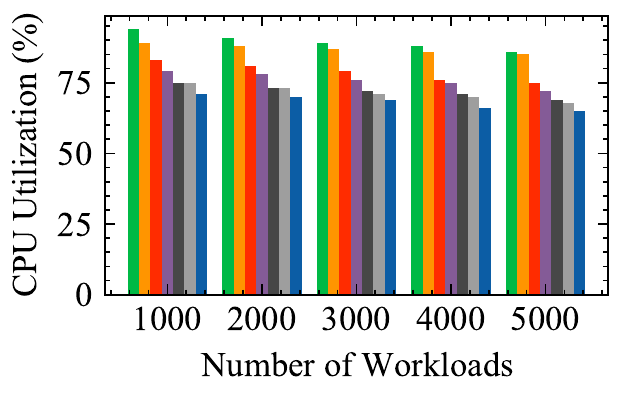}
    \label{fig:5f}
    }
    \subfigure[]{
    \includegraphics[width=.23\textwidth]{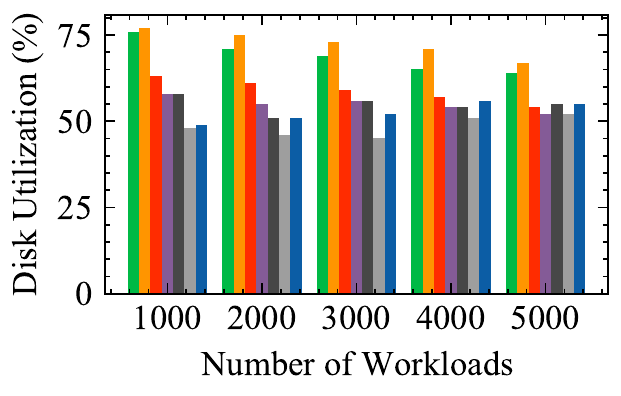}
    \label{fig:5g}
    }
    \subfigure[]{
    \includegraphics[width=.23\textwidth]{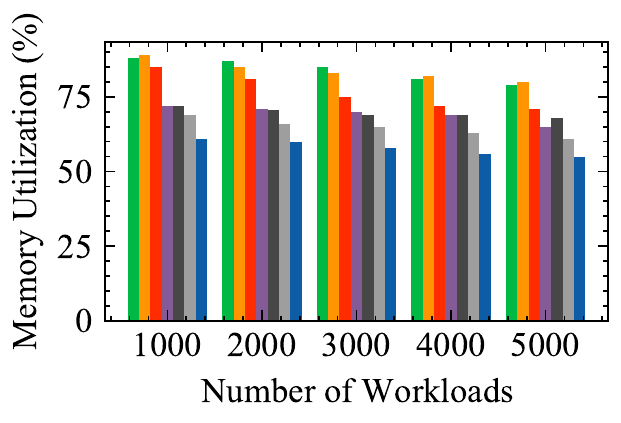}
    \label{fig:5h}
    }\\
    \caption{Comparison of performance parameters with different value of workloads: a) Execution Time, b) Resource Contention, c) Energy Consumption, d) SLA Violation Rate, e) Network Utilization, f) CPU Utilization, g) Disk Utilization and h) Memory Utilization }
    \label{fig:comparison3}
}
\end{figure*}

\subsubsection{Variation of Number of Workloads}

In this section we evaluate the value of various performance parameters as we increase the number of workloads.

Figure \ref{fig:5a} shows the variation of execution time with different numbers of workloads. The value of execution time in START is 19.74-23.84\% less than the baseline methods. The interpretation of resource contention for different numbers of workloads is shown in Figure \ref{fig:5b} which shows the value of resource contention increases with the increase in the number of workloads. START performs better than existing techniques; the average value of resource contention in START is 19.12-24.84\% less than the baseline methods. Figure \ref{fig:5c} shows the variation of energy consumption with different numbers of workloads and the value of energy consumption in START is 13.71-18.01\% less than the baseline methods. The variation of SLA violation rate for different number of workloads is shown in Figure \ref{fig:5d} and the value of SLA violation rate is increasing with the increase in number of workloads but START performs better than existing techniques. The average value of resource contention in START is 9.26-12.92\% less than the baseline methods. The reduced execution times (and hence energy consumption and SLA violations) are due to efficient and proactive mitigation of stragglers by START. Further, using the Pareto distribution allows START to identify stragglers prior to their completion, which reduces resource usage and hence contention.

 \begin{figure*}[!]
{
    \centering
    \includegraphics[width=.6\textwidth]{graphs/legend.pdf}
    \\
    \subfigure[]{
    \includegraphics[width=.115\textwidth]{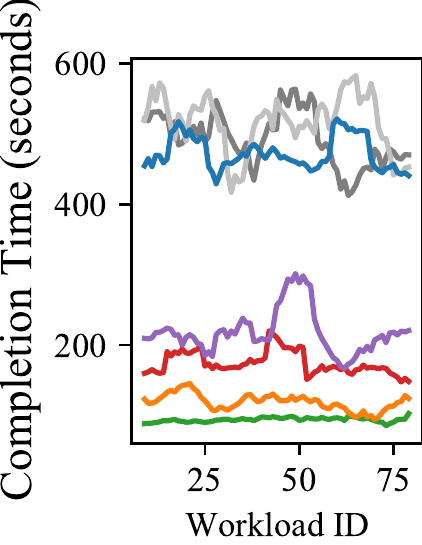}
    \includegraphics[width=.115\textwidth]{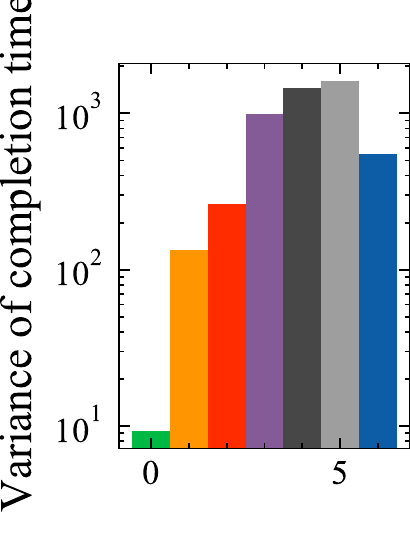}
    \label{fig:7a}
    }
    \subfigure[]{
    \includegraphics[width=.115\textwidth]{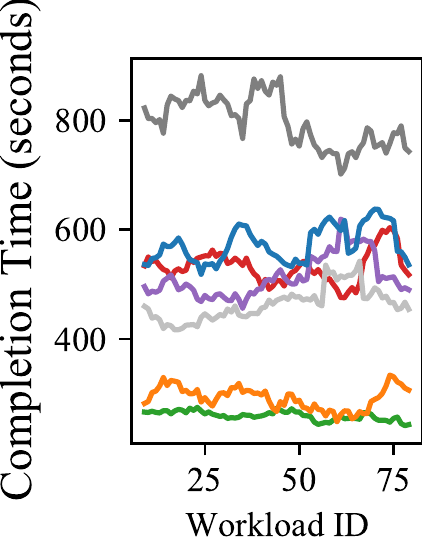}
    \includegraphics[width=.115\textwidth]{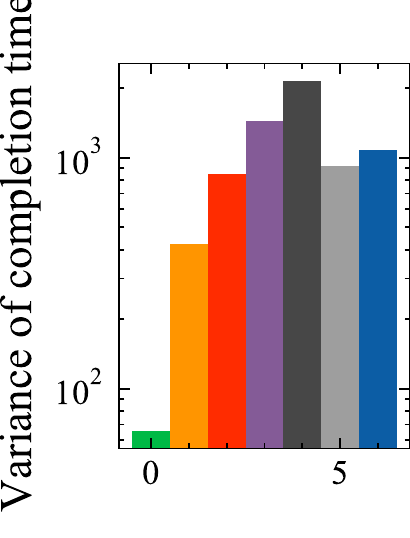}
    \label{fig:7b}
    }
    \subfigure[]{
    \includegraphics[width=.115\textwidth]{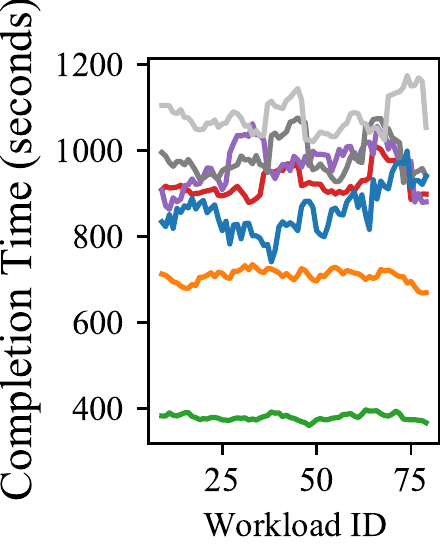}
    \includegraphics[width=.115\textwidth]{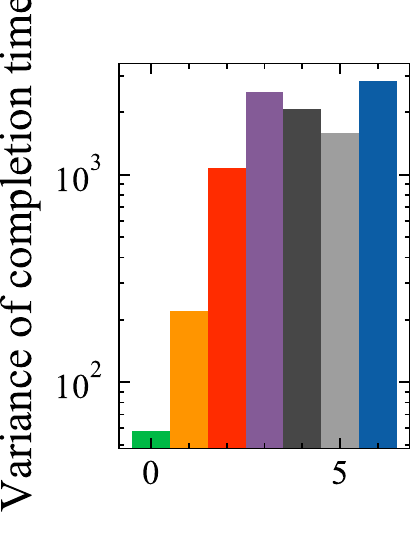}
    \label{fig:7c}
    }
    \subfigure[]{
    \includegraphics[width=.115\textwidth]{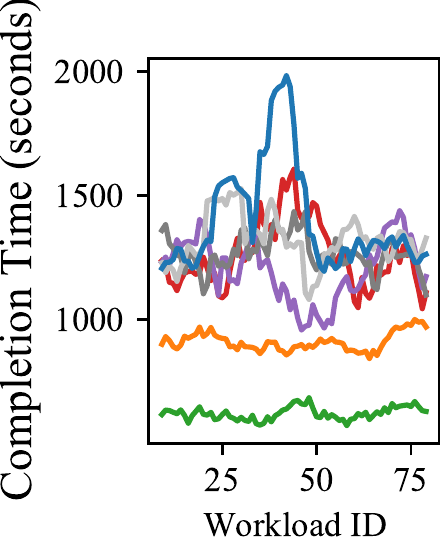}
    \includegraphics[width=.115\textwidth]{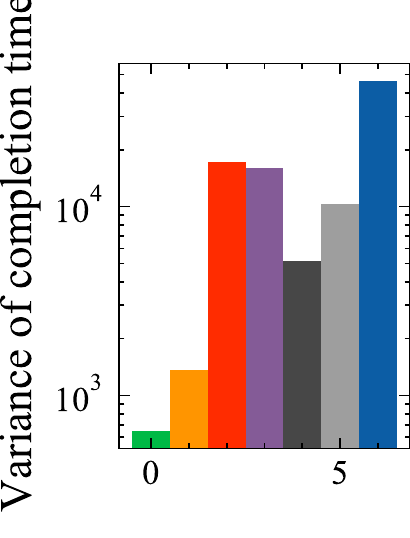}
    \label{fig:7d}
    }
    \caption{Comparison of performance based on execution time for different utilization: a) utilization limit = 20\%, b) utilization limit = 40\%, c) utilization limit = 60\% and d) utilization limit = 80\%}
    \label{fig:comparison3}
}
\end{figure*} 

Figure \ref{fig:5e} shows that the variation of network utilization with a different number of workloads for START and the baseline methods. \blue{All the utilization metrics presented in the figure are averaged across the completed tasks.} The experimental results show that the average value of network utilization in START is between 18.6\% and 25.67\% more than the baseline methods. The variation of CPU utilization with different numbers of workloads is shown in Figure \ref{fig:5f} and it shows the value of CPU utilization is decreasing with the increase in the number of workloads but START performs better than existing techniques. The value of CPU utilization in START is between 16.61\% and 17.29\% more than the baseline methods. Figure \ref{fig:5g} shows the variation of disk utilization with a different number of workloads for all methods. The experimental result show that the average value of disk utilization in START is 13.25-15.34\% more than the baseline methods. The variation of memory utilization with a different number of workloads is shown in Figure \ref{fig:5h} and indicates that the value of memory utilization is decreasing with the increase in the number of workloads but START performs better than existing techniques. The value of memory utilization in START is 7.92-17.54\% more than the baseline methods. The reduction in usage of resources in case of START is because of the conservative execution of tasks based on straggler prediction. Instead of running/speculating straggler tasks in advance, START waits for the completion of $q-\floor{E_s}$ (refer Algorithm~\ref{alg:scheduling}). Thus, if the predicted straggler tasks do complete earlier than expected, they are not cloned, avoiding resource wastage.

\subsection{Straggler Analysis}

Figure \ref{fig:comparison3} shows the \blue{variation of completion time of different workloads for different straggler management techniques with different utilization percentages of CPU, disk, memory and network. The line plots show the completion time across the workloads sorted by their creation time and the bar plots show the variation in the completion time.  A higher variance of completion time implies a higher number of tasks that cause a delay in job completion. Thus, a simple measure for comparison is the variance of execution times across different tasks. Figures \ref{fig:7a}, \ref{fig:7b}, \ref{fig:7c} and \ref{fig:7d} show the comparison of START with existing straggler management techniques for 20\%, 40\%, 60\% and 80\% reserved utilization respectively. The observed improvement occurs because START is very effective in the detection and mitigation of stragglers at run-time. It is also identified that the completion time is increasing with the increase in utilization limit from 20\% to 80\%}. Figure \ref{fig:7d} shows that START has more variation in job completion time with an 80\% utilization limit, but START performs better than existing techniques while detecting and mitigating stragglers more efficiently.

\subsection{Prediction Accuracy Comparison}

\begin{figure}[!t]
{
    \centering
    \begin{minipage}{0.6\columnwidth}
    \centering
        \includegraphics[width=\columnwidth]{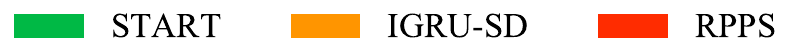} \\
        \subfigure[]{
        \includegraphics[width=\columnwidth]{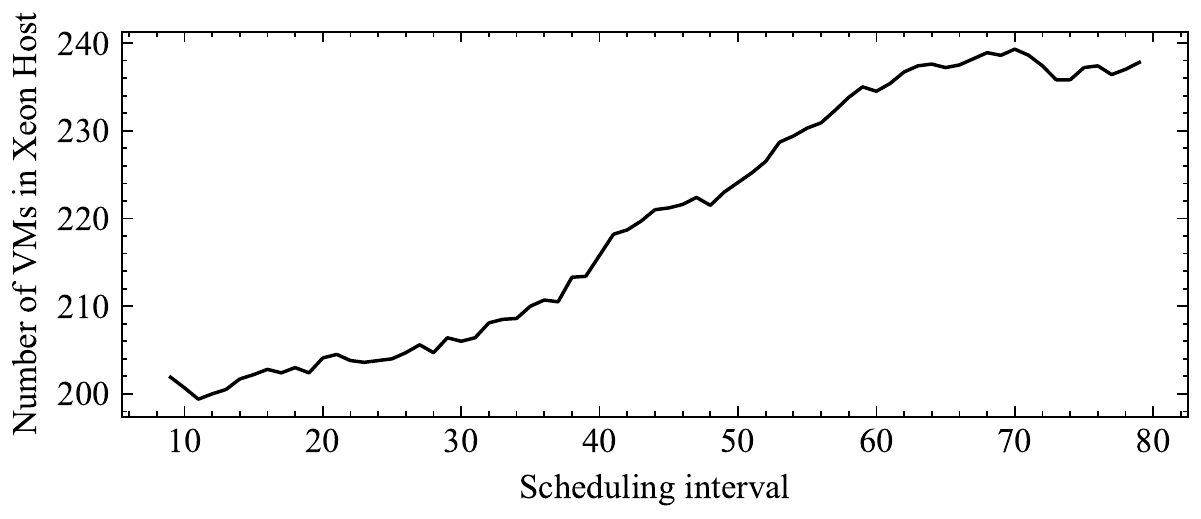}
        \label{fig:9a}
        }
        \includegraphics[width=\columnwidth]{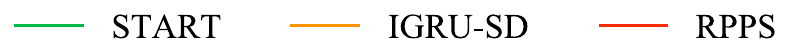}
        \\
        \subfigure[]{
        \includegraphics[width=\columnwidth]{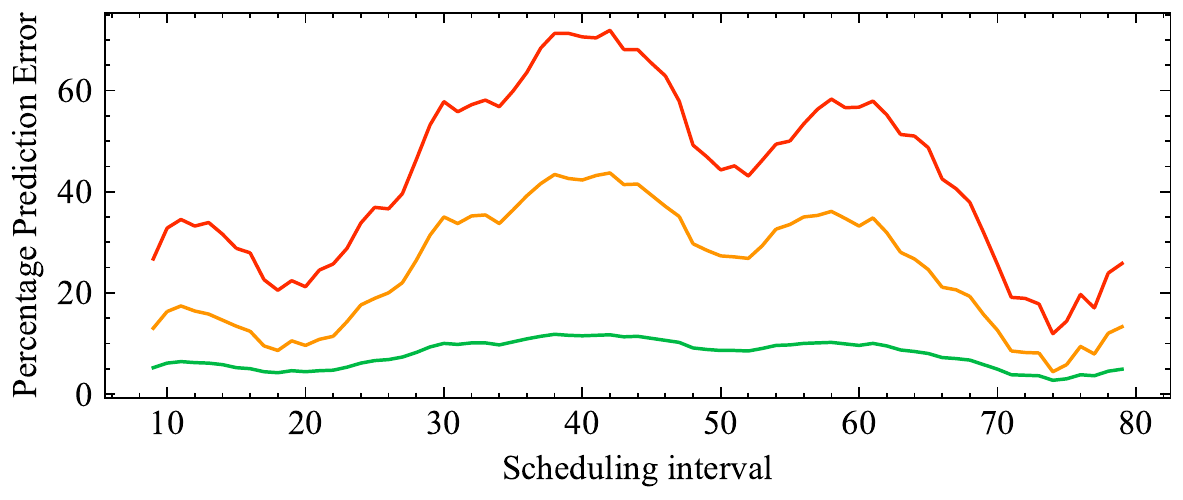}
        \label{fig:9b}
        }
    \end{minipage}
    \begin{minipage}{0.35\columnwidth}
    \centering
        \subfigure[]{
        \includegraphics[width=0.6\columnwidth]{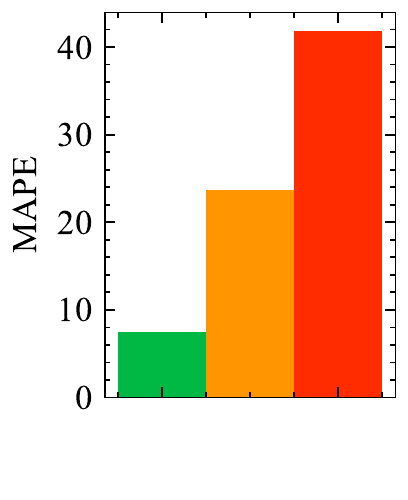} 
        \label{fig:9c}
        }
        \\
        \subfigure[]{
        \includegraphics[width=0.6\columnwidth]{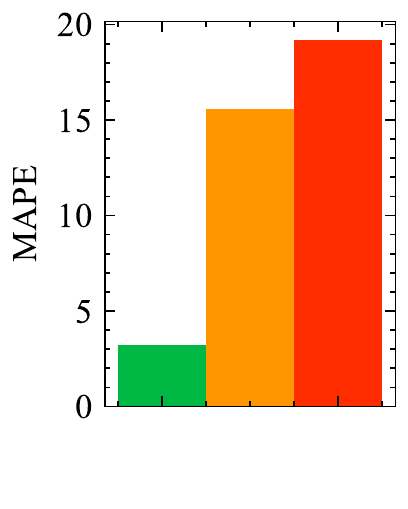}
        \label{fig:9d}
        }
    \end{minipage}
    \caption{Comparison of prediction accuracy of START with IGRU-ISD and RPPS. (a) Number of VMs in Xeon host out of total 400 VMs, (b) Comparison of percentage prediction error, (c) MAPE values for modified environment with changing host resources (d) MAPE values for initial setup described in Section \ref{sec:exp}.}
    \label{fig:accuracy}
}
\end{figure}

To demonstrate the efficacy of the prediction model, we show that the prediction error is minimized in our model. To evaluate prediction error, we use the same environment as before with diverse task requirements and heterogeneous hosts with host failures. We use the MAPE metric for this. For ease of comparison, we consider only 2 physical host types with processors: i5 and Xeon as given in Table \ref{tab:setup}. We keep a total 200 VMs out of which the number of VMs on the Xeon host are changed with time (the variation is not smooth due to injected VM failures in the model). As shown in Figure \ref{fig:accuracy}, as the number of VMs on the Xeon host change, the percentage prediction error is higher for RPPS and IGRU-SD than START. This is because these models do not consider the heterogeneity of VM resource capabilities. Clearly, when the number of VMs in the Xeon host change, the heterogeneity changes dynamically, leading to different probabilities of tasks becoming stragglers. Thus, the models in IGRU-SD and RPPS are unable to predict straggler tasks accurately. In contrast, START is able to analyze host resource capabilities with the task allocation to correctly predict straggler tasks.

\subsection{Overhead Comparison}

\begin{figure}[!t]
    \centering \setlength{\belowcaptionskip}{-10pt}
    \includegraphics[width=\columnwidth]{graphs/legend.pdf}\\
    \includegraphics[width=0.75\columnwidth]{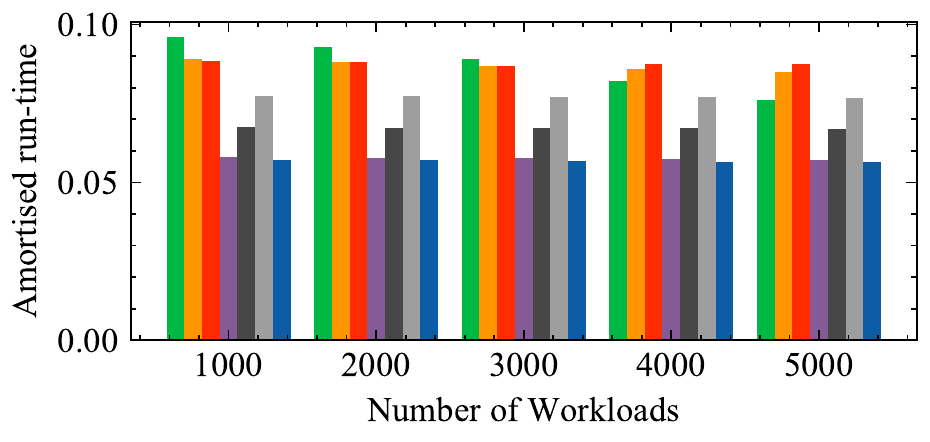}
    \caption{Overhead comparison}
    \label{fig:overhead}
\end{figure}

Figure~\ref{fig:overhead} shows a comparison of run-times of the START and baseline approaches (including scheduling of re-run or speculated tasks) amortized over the average task execution times. As can be seen, the methods proposed in the prior art are faster at detecting straggler tasks. However, as seen earlier, they do not perform well. START has a slightly higher ($0.09\%$) run-time than the best approach among the prior work (IGRU-SD).  

\section{Conclusions and Future Work}
\label{sec:conclusion}

We proposed a novel straggler prediction and mitigation technique using an Encoder-LSTM Model for large-scale cloud computing environments. This technique allows us to reduce response time and provide better results with fewer SLA violations compared to prior works. Thanks to the prediction models based on maximum likelihood estimation from a Pareto distribution and recurrent encoder network, our model is able to predict straggler tasks beforehand and mitigate them early on using speculation and re-run methods. Unlike prior prediction based approaches, START is able to analyze tasks with host characteristics and utilize the underlying Pareto distribution for more accurate prediction and mitigation leading to higher performance than state-of-the-art mechanisms. It is clear that for different workload levels, START performs better giving lower execution time, resource contentions, energy consumption and SLA violation rate. When compared with different levels of workload on the cloud system, again START outperforms the baseline approaches. 
START has higher CPU, network, RAM and disk utilization. This is because many jobs, and hence, tasks complete quickly which leads to more tasks being finished in a period of time compared to other approaches. This implies that START is able to leverage resources in a more efficient manner leading to faster job completion and hence also saving energy, even with slightly higher resource utilization for the same number of tasks.

As part of future work, we plan to implement START in real-life settings using fog frameworks such as PRISM~\cite{lindsay2019prism} or COSCO~\cite{tuli2021cosco}. This will help in making the model more robust to task and workload stochasticity in real scenarios. Moreover, we can also fine-tune our neural network models and Pareto distribution parameters using a larger dataset which includes diverse fog and cloud applications. 

\section*{Acknowledgements}
\footnotesize{
S.T. is grateful to the Imperial College London for funding his Ph.D. through the President’s Ph.D. Scholarship scheme. P.G. and S.S.G are supported by the Engineering and Physical Sciences Research Council (EPSRC) (EP/P031617/1). R.B. is supported by Melbourne-Chindia Cloud Computing (MC3) Research Network and Australian Research Council. The work of G.C. has been partly funded by the EU’s Horizon 2020 program under grant agreement No 825040.}

\bibliographystyle{IEEEtran}

\bibliography{references}
\vspace{-0.5in}
\begin{IEEEbiography}
[{\includegraphics[width=1in,height=1in,clip,keepaspectratio]{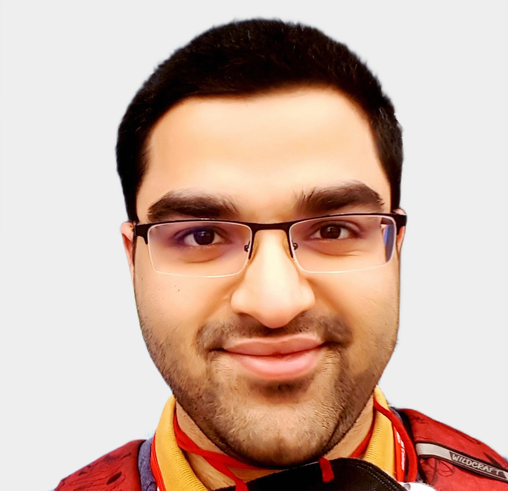}}]
{Shreshth Tuli}
is a President's Ph.D. Scholar at the Department of Computing, Imperial College London, UK. Prior to this he was an undergraduate student at the Department of Computer Science and Engineering at Indian Institute of Technology - Delhi, India. He has worked as a visiting research fellow at the CLOUDS Laboratory, School of Computing and Information Systems, the University of Melbourne, Australia. His research interests include Internet of Things, Fog Computing and Deep Learning.
\end{IEEEbiography}
\vspace{-0.45in}
\begin{IEEEbiography}
[{\includegraphics[width=1in,height=1in,clip,keepaspectratio]{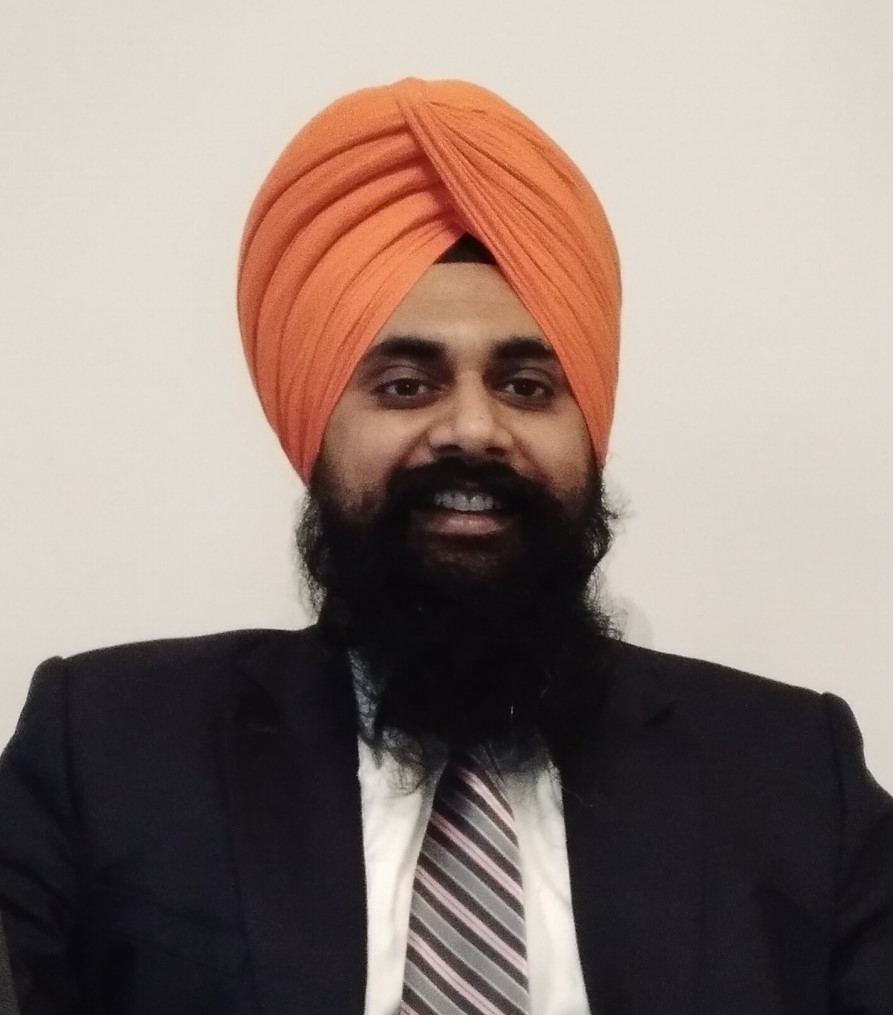}}]
{Sukhpal Singh Gill}
is a Lecturer (Assistant Professor) in Cloud Computing at School of EECS, Queen Mary University of London, UK. Prior to this, Dr. Gill has held positions as a Research Associate at the School of Computing and Communications, Lancaster University, UK and also as a Postdoctoral Research Fellow at CLOUDS Laboratory, The University of Melbourne, Australia. His research interests include Cloud Computing, Fog Computing, Software Engineering, Internet of Things and Big Data. 
\end{IEEEbiography}
\vspace{-0.4in}
\begin{IEEEbiography}
 [{\includegraphics[width=1in,height=1in,clip,keepaspectratio]{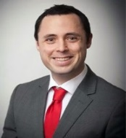}}]
    {Peter Garraghan }
is a Reader in the School of Computing \& Communications, Lancaster University. His primary research expertise is studying the complexity and emergent behaviour of massive-scale distributed systems (Cloud computing and Internet of Things) to propose design new techniques for enhancing system dependability, resource management, and energy-efficiency. He has collaborated internationally with the likes of Alibaba Group and Microsoft.
\end{IEEEbiography}
\vspace{-0.45in} 
\begin{IEEEbiography}
  [{\includegraphics[width=1in,height=1in,clip,keepaspectratio]{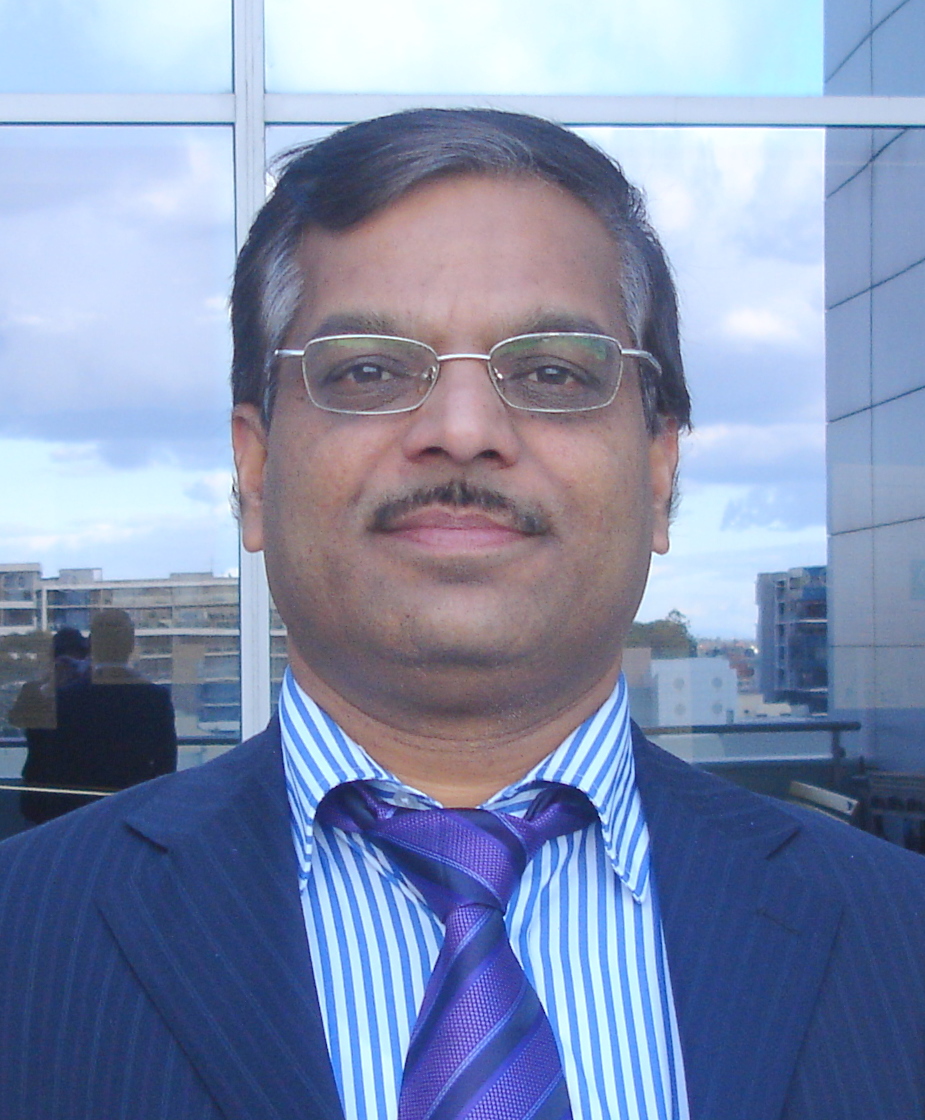}}]
    {Rajkumar Buyya} is a Redmond Barry Distinguished Professor and Director of the Cloud Computing and Distributed Systems (CLOUDS) Laboratory at the University of Melbourne, Australia. He has authored over 625 publications and seven textbooks including "Mastering Cloud Computing" published by McGraw Hill, China Machine Press, and Morgan Kaufmann for Indian, Chinese and international markets respectively. He is one of the highly cited authors in computer science and software engineering worldwide (h-index=150, g-index=322, 117,000+ citations).  He is a fellow of the IEEE.
\end{IEEEbiography}
\vspace{-0.45in}
\begin{IEEEbiography}
[{\includegraphics[width=1in,height=1in,clip,keepaspectratio]{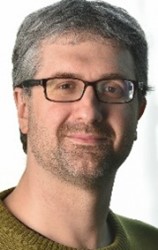}}]
{Giuliano Casale}
joined the Department of Computing at Imperial College London in 2010, where he is currently a Reader. He teaches and does research in performance engineering and cloud computing, topics on which he has published more than 100 refereed papers. He has served for several conferences in the area of performance and reliability engineering, such as ACM SIGMETRICS/Performance and IEEE/IFIP DSN. His research work has received multiple awards, recently the best paper award at ACM SIGMETRICS. He serves on the editorial boards of IEEE TNSM and ACM TOMPECS and as current chair of ACM SIGMETRICS.
\end{IEEEbiography}
\vspace{-0.4in}
\begin{IEEEbiography}
[{\includegraphics[width=1in,height=1in,clip,keepaspectratio]{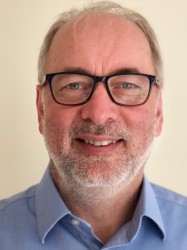}}]
{Nicholas R. Jennings}
is the Vice-Chancellor and President of Loughborough University. He is an internationally-recognised authority in the areas of AI, autonomous systems, cyber-security and agent-based computing. He is a member of the UK government’s AI Council, the governing body of the Engineering and Physical Sciences Research Council, and chair of the Royal Academy of Engineering’s Policy Committee.  Before Loughborough, he was the Vice-Provost for Research and Enterprise and Professor of Artificial Intelligence at Imperial College London, the UK's first Regius Professor of Computer Science (a post bestowed by the monarch to recognise exceptionally high quality research) and the UK Government’s first Chief Scientific Advisor for National Security.
\end{IEEEbiography}
\vfill

\end{document}